\documentclass[superscriptaddress,aalso receivedpaper,10pt,twocolumn,floats,showpacs,amsmath,amsfonts,amssymb,reprint]{revtex4-2}

\usepackage{graphicx}
\usepackage[colorlinks=true,citecolor=blue]{hyperref}
\usepackage[caption=false]{subfig}
\usepackage{pstricks}
\usepackage{pst-node}
\usepackage{amsmath}
\usepackage{amsbsy}
\usepackage{amsfonts}
\usepackage{verbatim}
\usepackage{dsfont}
\usepackage{MnSymbol}
\usepackage{graphicx,color}
\usepackage[table]{xcolor}

\newcommand{\re}{\text{Re}}

\newcommand{\Tr}{\text{Tr}}

\begin{document}

	\title{Quantum instanton theory for metastable collective spin systems}
\author{Krzysztof Ptaszy\'{n}ski}
	\email{krzysztof.ptaszynski@ifmpan.edu.pl}
 	\affiliation{Institute of Molecular Physics, Polish Academy of Sciences, Mariana Smoluchowskiego 17, 60-179 Pozna\'{n}, Poland}

\author{Maciej Chudak}
 	\affiliation{Institute of Molecular Physics, Polish Academy of Sciences, Mariana Smoluchowskiego 17, 60-179 Pozna\'{n}, Poland}
    
	\author{Massimiliano Esposito}
\affiliation{Complex Systems and Statistical Mechanics, Department of Physics and Materials Science, University of Luxembourg, 30 Avenue des Hauts-Fourneaux, L-4362 Esch-sur-Alzette, Luxembourg}
	
	\date{\today}

\begin{abstract}
Collective spin systems---driven-dissipative spin ensembles whose unitary and dissipative processes act uniformly on all spins, allowing their dynamics to be described as a single macrospin---play an important role in atomic and solid-state physics. Their intrinsic nonlinearity gives rise to multiple long-lived metastable states that ultimately relax to a unique most probable state. This dominant state can change with a control parameter, leading to first-order phase transitions. We develop a real-time instanton approach based on quantum quasiprobability dynamics that captures the stationary state in the large-spin limit and the asymptotic scaling of relaxation rates. We further show that these features are not accurately described by the previously applied semiclassical Wigner approach due to its neglect of non-Gaussian fluctuations.
\end{abstract}

\maketitle

\textit{Introduction}---Driven–dissipative quantum systems can exhibit multistability, i.e., the existence of several long-lived metastable states~\cite{macieszczak2021theory,carr2013nonequilibrium,rodriguez2017probing,fink2018signatures,chen2023quantum,beaulieu2025observation,noachtar2026exact}. This typically arises when a system admits a deterministic mean-field (MF) description as some extensivity parameter $V$ (e.g., the number of atoms in atomic ensembles or the inverse nonlinearity strength in quantum resonators) tends to infinity. Then, metastable states correspond to infinitely long-lived attractors of the MF dynamics. 
For finite $V$, rare fluctuations induce transitions between attractors, giving them finite lifetimes~\cite{drummond1989quantum,kinsler1991quantum,lee2012colllective}---an effect observed experimentally in quantum resonators~\cite{rodriguez2017probing,fink2018signatures,chen2023quantum,beaulieu2025observation}. The resulting switching can be modeled as a classical Markov jump process between attractors~\cite{macieszczak2021theory}. Except in fine-tuned cases, this process relaxes the system to a single most probable attractor. As some control parameter varies, one of the attractors may become most probable at the expense of the other, causing sharp jumps in observables that become increasingly abrupt with $V$, signaling a first-order (discontinuous) dissipative phase transition~\cite{minganti2018spectral,chen2023quantum,beaulieu2025observation,roberts2023exact}.

The analysis of metastability using methods such as the quantum master equation (QME) becomes increasingly challenging at large $V$. An effective approach is to exploit Arrhenius-like asymptotic scaling of switching rates between attractors
$i$ and $j$~\cite{dykman1988quantum,dykman2007critical,drummond1989quantum,kinsler1991quantum,lee2024real,xiang2025switching}
\begin{align} \label{eq:arrh}
\kappa_{i\rightarrow j} \asymp \exp(-V\mathcal{A}_{i \rightarrow j}) \,,
\end{align}
where 
$a \asymp b$ denotes the exponential asymptotics $\lim_{V \rightarrow \infty} (\ln a)/(\ln b)=1$. The quantities $\mathcal{A}_{i \rightarrow j}$ are \textit{activation barriers}, which govern the steady state (most probable attractor) and long-time dynamics of the system. As proposed already long ago~\cite{dykman1988quantum,dykman2007critical}, for bosonic quantum resonators, these barriers can be computed via a real-time instanton approach, as the action of an auxiliary dynamical system. This method, usually based on the Keldysh path integral formalism, has recently gained prominence~\cite{lee2024real,sepulcre2026analytical,mylnikov2026large}, e.g., in describing bit-flip errors in Schr\"{o}dinger cat qubits~\cite{thompson2022qubit,carde2026nonperturvative,mylnikov2025switching,mylnikov2025qubit,thompson2026spectroscopy}.

Here, we extend the instanton approach to describe metastability in collective spin systems---ensembles of $N$ spins in which both unitary and dissipative processes act uniformly on all constituents, rendering the dynamics invariant under spin permutations. Their many-body dynamics can then be mapped onto a single collective degree of freedom---a macrospin (whose length may vary under local dissipation) represented by the collective spin operators $\hat J_{x,y,z}=\sum_{i=1}^N\sigma_{x,y,z}^i/2$, where $\sigma_{x,y,z}^i$ denotes the local Pauli operator of the $i$th spin. Such models naturally arise in cavity or circuit quantum electrodynamics on atomic~\cite{mivehvar2021cavity,morrison2008dynamicalprl,morrison2008collective,norcia2018cavity,muniz2020exploring,song2025dissipation,ferri2021emerging} and solid-state platforms~\cite{mlynek2012demonstrating,nissen2013collective}, and also describe coupled electron and nuclear spins~\cite{kessler2012dissipative} and atoms on surfaces~\cite{shakirov2016role,ferreira2019lipkin}.

Although Ref.~\cite{lee2024real} proposed that metastability in such systems could be described by extending the Keldysh path-integral approach developed for bosonic systems, this has not yet been demonstrated explicitly. Moreover, path-integral formulations for spin systems are known to be technically involved~\cite{kiselev2000schwinger}. Consequently---apart from direct QME simulations---previous studies of metastability in spin systems 
have relied on semiclassical Fokker--Planck equations for the Wigner distribution~\cite{dutta2025quantum,xiang2025switching}. However, as we demonstrate, such approaches can fail to accurately capture activation barriers due to the neglect of non-Gaussian fluctuations. 
In this work, we overcome these limitations by constructing an instanton approach from the exact, nontruncated equations of motion for quantum quasiprobability distributions. The use of nontruncated equations has been previously explored for bosonic systems but was restricted to systems where they can be solved analytically~\cite{drummond1989quantum,kinsler1991quantum,mylnikov2025qubit}. Our framework builds on the observation that quantum quasiprobability dynamics, while exhibiting important differences, shares a structural resemblance with classical stochastic processes, where instanton methods are well established~\cite{meier1993escape,dykman1994large,grafke2019numerical,gagrani2023action,zakine2023minimum,FalascoReview}. This enables us to adapt techniques from the classical setting to accurately characterize activation barriers---and thereby the steady-state and relaxation timescale---in multistable collective spin systems.

\textit{Setup---}We illustrate our formalism on a concrete example of the minimal bistable collective spin model introduced in Ref.~\cite{dutta2025quantum}. Although not derived from a microscopic physical model, its simplifying symmetry, discussed below, makes it a particularly transparent setting for illustrating the basic qualitative features of the instanton formalism. Applications of our framework to microscopically motivated and experimentally relevant models are presented in the companion paper~\cite{chudak2026quantum}.

The model obeys the Lindblad QME ($\hbar=1$)
\begin{align}
\label{eq}
\partial_t \hat{\rho}
=\mathcal{L}\hat{\rho}
\equiv
-i\left[\Omega \hat J_x,\hat{\rho}\right]
+\frac{\gamma}{J}\mathcal{D}[\hat J_+]\hat{\rho}
+\frac{\Gamma}{J^3}\mathcal{D}[\hat J_-\hat J_z]\hat{\rho},
\end{align}
where $\hat{\rho}$ is the system density matrix, $\mathcal{L}$ is the Liouvillian superoperator, $\mathcal{D}[\hat A]\bullet\equiv \hat A\bullet\hat A^\dagger-\{\hat A^\dagger\hat A,\bullet\}/2$ is the dissipator, $\hat J_\pm\equiv \hat J_x\pm i\hat J_y$ are collective ladder operators, and $\Omega$, $\gamma$, and $\Gamma$ denote the strengths of coherent driving, linear pumping, and nonlinear dissipation, respectively. Compared with Ref.~\cite{dutta2025quantum}, we include the coherent drive $\Omega$, thereby precluding a description of the system dynamics as a Markov jump process among the eigenstates of $\hat J_z$ and thus making it intrinsically quantum. 

In this model, the Hamiltonian and jump operators involve only collective spin operators $\hat J_k$ and hence commute with $\hat{\mathbf J}^2=\hat J_x^2+\hat J_y^2+\hat J_z^2$. Consequently, the total spin angular momentum $J$, defined by the eigenvalue $J(J+1)$ of $\hat{\mathbf J}^2$, is conserved by the dynamics. This corresponds to strong permutation symmetry. The companion paper~\cite{chudak2026quantum} also treats an experimentally relevant model with local dissipation and weak permutation symmetry~\cite{song2025dissipation}, where $J$ is not conserved~\cite{shammah2018open}.

\begin{figure}
    \centering
    \includegraphics[width=0.9\linewidth]{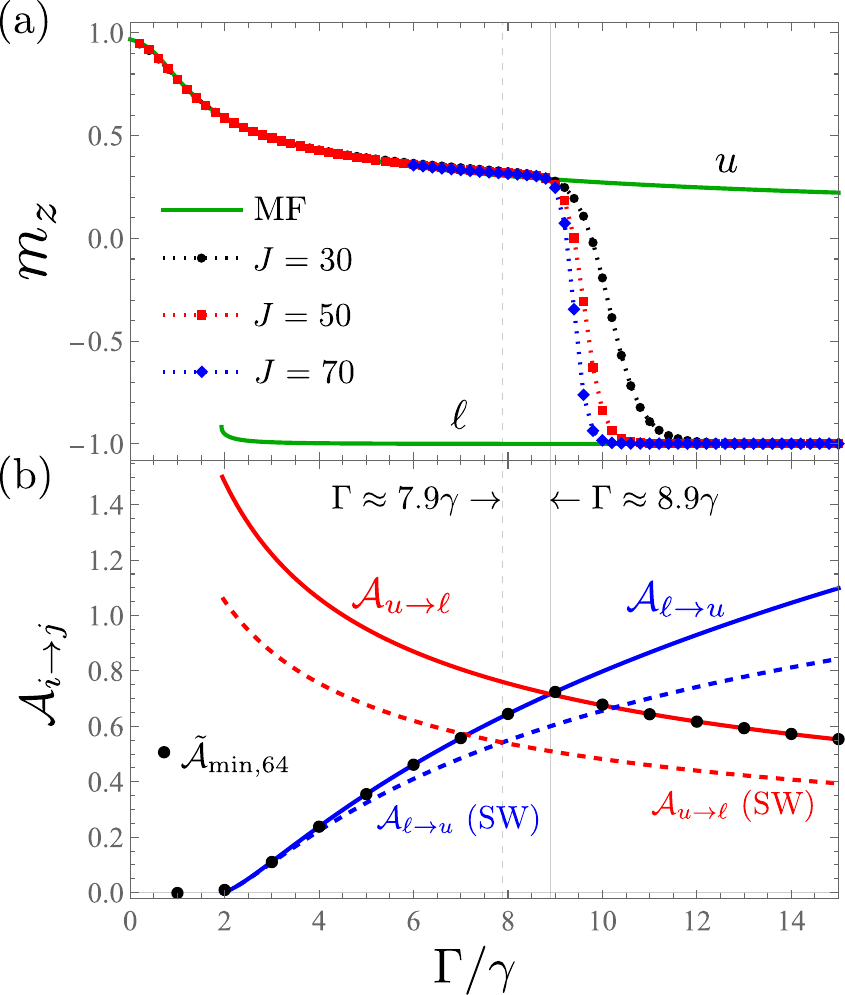}
    \caption{(a) MF fixed-point magnetization $m_z^*$ (green solid lines) versus finite-size steady-state magnetization $m_{z,\text{ss}}$ obtained using QME (symbols). QME calculation details in~\cite{supp}. (b) Activation barriers $\mathcal{A}_{i \rightarrow j}$ calculated using our approach (solid lines) versus SW approach (dashed lines). Grey vertical solid (dashed) line denotes the crossing point of $\mathcal{A}_{\ell \rightarrow u}$ and $\mathcal{A}_{u \rightarrow \ell}$ curves for our (SW) approach. Dots denote the estimator $\tilde{\mathcal{A}}_{\text{min},64}$ defined by Eq.~\eqref{eq:aminest}. Parameter $\Omega=0.25 \gamma$.
    }
    \label{fig:mzaction}
\end{figure}

In the limit of $J \rightarrow \infty$ the model is exactly described by nonlinear MF equations for the magnetization components $m_k \equiv\Tr[\hat{\rho} \hat{J}_k]/J$ (see Appendix~\ref{app:mf}),
 \begin{align} \nonumber
d_t m_x&=(\Gamma m_z^2-\gamma) m_x m_z \,, \\ \label{eq:meanfield}
d_t m_y &=-\Omega m_z+(\Gamma m_z^2-\gamma) m_y m_z \,, \\ \nonumber
d_t m_z &=\Omega m_y-(\Gamma m_z^2-\gamma) (m_x^2+m_y^2) \,,
\end{align}
with $m_x^2+m_y^2+m_z^2=1$. We further focus on the case of moderate drive $\Omega=0.25\gamma$, where the MF equations relax the system to stable fixed points (FPs) $m_k=m_k^*$ for which $d_t m_k=0$; see the stability analysis in Appendix~\ref{app:stability}. The magnetization component $m_z^*$ associated with these FPs is presented in Fig.~\ref{fig:mzaction}(a).  For small $\Gamma \lessapprox 1.94 \gamma$, the system has a unique stable FP with $m_z^*>0$, which we call the upper branch (denoted $u$). For $\Gamma \gtrapprox 1.94 \gamma$, this FP is still stable, but a second stable FP with $m_z^* \approx -1$ emerges, which we call the lower branch (denoted $\ell$). This stands in contrast to the QME approach, which---for finite $J$---always admits a unique stationary state $\mathcal{L} \hat{\rho}_\text{ss}=0$. The corresponding magnetization value $m_{z,\text{ss}} \equiv \Tr[\hat{\rho}_\text{ss} \hat{J}_z]/J$ is denoted by symbols in Fig.~\ref{fig:mzaction}(a): for small $\Gamma \lessapprox 8.9 \gamma$ it focuses around the $u$ branch of the MF solution, but at $\Gamma \approx 8.9 \gamma$, it suddenly deviates from it and decays towards the $\ell$ branch. The larger $J$, the faster the decay.

This apparent incongruence of the MF and QME approaches is known as Keizer's paradox~\cite{keizer1978thermodynamics,FalascoReview}. It results from the fact that, for finite $J$, the MF attractors convert to metastable states with long but finite lifetimes. Since $J$ is the extensive parameter $V$ here, from Eq.~\eqref{eq:arrh}, we find that the ratio of probabilities of the $u$ and $\ell$ branches scales as $P_u/P_\ell=\kappa_{\ell \rightarrow u}/\kappa_{u \rightarrow\ell} \asymp \exp[J (\mathcal{A}_{u \rightarrow \ell}-\mathcal{A}_{\ell \rightarrow u})]$. Thus, the magnetization behavior in Fig.~\ref{fig:mzaction}(a) suggests that for $\Gamma \lessapprox 8.9 \gamma$, $\mathcal{A}_{u \rightarrow \ell}>\mathcal{A}_{\ell \rightarrow u}$, making the escape from the $u$ branch exponentially more difficult as $J$ increases, and thus its occupation tends to one. The situation is reversed for $\Gamma \gtrapprox 8.9 \gamma$. In the limit $J \rightarrow \infty$ (taken after the $t \rightarrow \infty$ limit), one thus expects that the system exhibits a first-order dissipative phase transition at the point where $\mathcal{A}_{u \rightarrow \ell}=\mathcal{A}_{\ell \rightarrow u}$, with $m_{z,\text{ss}}$ jumping discontinuously from the $u$ to the $\ell$ branch.

\textit{Instanton approach}---We now present a method to determine the activation barriers $\mathcal{A}_{i \rightarrow j}$ without solving the QME, which becomes cumbersome for large $J$. To do so, we first represent the density matrix using the Husimi and P representations $p_H$ and $p_P$ defined as~\cite{narducci1975multitime,altland2012quantum,mandt2015stochastic,burkle2020probabilistic}
\begin{subequations}
\begin{align}
&p_H(\theta,\varphi)  \equiv \tfrac{2J+1}{4\pi} \langle \theta,\varphi |\hat{\rho} |\theta,\varphi \rangle \,, \\
&\hat{\rho} \equiv \int  p_P(\theta,\varphi) |\theta,\varphi \rangle \langle \theta,\varphi |d\theta d\varphi \,,
\end{align}
\end{subequations}
where $|\theta,\varphi \rangle$ are coherent spin states~\cite{radcliffe1971some}
\begin{align}
&|\theta,\varphi \rangle \equiv \\\nonumber  &\sum_{M=-J}^J \sqrt{2J \choose J+M} \left( \cos\frac{\theta}{2} \right)^{J+M} \left( \sin\frac{\theta}{2} e^{i \varphi} \right)^{J-M} |J,M \rangle \,,
\end{align}
and $|J,M \rangle$ are eigenstates of $\hat{J}_z |J,M \rangle=M|J,M \rangle$. Moving from spherical coordinates $\theta$, $\varphi$ to stereographic coordinates $v$, $w$ that fulfill $\cot(\theta/2)e^{i \varphi}=v+iw$, the dynamics of both distributions, $p_{\alpha}$ where $\alpha \in \{H,P \}$, are given by 
\begin{align} \label{eq:evquas}
\partial_t p_\alpha (\boldsymbol{x},t)=L_\alpha(\boldsymbol{x},\nabla_{\boldsymbol{x}}) p_\alpha(\boldsymbol{x},t) \,,
\end{align}
where $L_\alpha(\boldsymbol{x},\nabla_{\boldsymbol{x}})$ is the differential operator representing $\mathcal{L}$, $\boldsymbol{x} \equiv (v,w)$, and $\nabla_{\boldsymbol{x}}$ is the gradient over $\boldsymbol{x}$. The method to derive $L_\alpha(\boldsymbol{x},\nabla_{\boldsymbol{x}})$ is discussed in Refs.~\cite{narducci1975multitime,altland2012quantum,mandt2015stochastic,burkle2020probabilistic} and briefly reviewed in Appendix~\ref{app:derevquas}. 
We further define the propagators $K_\alpha(\boldsymbol{x}'',t|\boldsymbol{x}',0)$ of these dynamics,
\begin{align}
p_\alpha(\boldsymbol{x}'',t) \equiv \int d \boldsymbol{x}' K_\alpha(\boldsymbol{x}'',t|\boldsymbol{x}',0) p_\alpha(\boldsymbol{x}',0)\;,
\end{align}
which also obey Eq.~\eqref{eq:evquas} and thus $\partial_t K_\alpha=L_\alpha K_\alpha$. As we show below, this propagator can be used to determine activation barriers $\mathcal{A}_{i \rightarrow j}$.
As follows from the expressions in Appendix~\ref{app:derevquas}, each $n$th-order derivative term of $L_\alpha(\boldsymbol{x},\nabla_{\boldsymbol{x}})$ scales as $J^{1-n}$. Consequently, the propagator can be represented using the WKB ansatz
\begin{align} \label{eq:wkb-prop}
K_\alpha(\boldsymbol{x}'',t|\boldsymbol{x'},0) \asymp e^{-J S_\alpha(\boldsymbol{x''},t|\boldsymbol{x}',0)} \,.
\end{align}
Substituting Eq.~\eqref{eq:wkb-prop} into $\partial_t K_\alpha=L_\alpha K_\alpha$ and taking the limit $J \rightarrow \infty$, one finds that $S_\alpha(\boldsymbol{x}'',t |\boldsymbol{x}',0)$ evolves according to the Hamilton--Jacobi equation~\cite{dykman1988quantum,FalascoReview}
\begin{align} \label{eq:hamilton-jacobi}
\partial_t S_\alpha(\boldsymbol{x}'',t|\boldsymbol{x}',0) =-\mathcal{H}_{\alpha}[\boldsymbol{x}'',\nabla_{\boldsymbol{x}''}S_\alpha(\boldsymbol{x}'',t|\boldsymbol{x}',0)] \,,
\end{align}
where
\begin{align} \label{eq:auxhamdef}
\mathcal{H}_\alpha(\boldsymbol{x},\boldsymbol{\pi})=\lim_{J \rightarrow \infty} J^{-1} L_\alpha(\boldsymbol{x},-J \boldsymbol{\pi}) \,,
\end{align}
is the auxiliary Hamiltonian with $\boldsymbol{\pi}\equiv(\pi_v,\pi_w)=\nabla_{\boldsymbol{x}''} S_\alpha$ playing the role of momentum. 
The solutions of Eq.~\eqref{eq:hamilton-jacobi} can be represented in terms of the action~\cite{dykman1988quantum,FalascoReview}
\begin{align} \nonumber
&S_\alpha(\boldsymbol{x}'',t|\boldsymbol{x'},0) = \\ \label{eq:action} &\int_0^t dt' \left\{ \boldsymbol{\pi}(t') \cdot d_{t'} \boldsymbol{x}(t')-\mathcal{H}_\alpha[\boldsymbol{x}(t'),\boldsymbol{\pi}(t')] \right\} \,,
\end{align}
evaluated along the \textit{instanton} trajectory $\boldsymbol{\xi}=[\boldsymbol{x}(t),\boldsymbol{\pi}(t)]$, obeying Hamilton's equations,
\begin{align} \label{eq:hamilton-eq}
d_t \boldsymbol{x}=\nabla_{\boldsymbol{\pi}} \mathcal{H}_\alpha (\boldsymbol{x},\boldsymbol{\pi}) \,, \quad d_t \boldsymbol{\pi}=-\nabla_{\boldsymbol{x}} \mathcal{H}_\alpha (\boldsymbol{x},\boldsymbol{\pi}) \,,
\end{align}
with the boundary conditions $\boldsymbol{x}(0)=\boldsymbol{x}'$ and $\boldsymbol{x}(t)=\boldsymbol{x}''$. 

We note that Eq.~\eqref{eq:action} can be multivalued, as $\boldsymbol{x}'$ and $\boldsymbol{x}''$ may be connected by different trajectories $\boldsymbol{\xi}$. For Hamiltonians $\mathcal{H}_\alpha$ that are convex in $\boldsymbol{\pi}$ (as for classical stochastic systems), the propagator is determined by the minimum action $S_\alpha$ among all solutions~\cite{FalascoReview,grafke2019numerical,gagrani2023action,zakine2023minimum}. The minimizing trajectories may change discontinuously as a function of $\boldsymbol{x}'$, $\boldsymbol{x}''$, producing nonanalytic behaviors of $S_\alpha$~\cite{bertini2010lagrangian}. 
In our case---as in frameworks employing Keldysh path integrals~\cite{lee2024real,carde2026nonperturvative}---the Hamiltonians $\mathcal{H}_\alpha$ are generically nonconvex in $\boldsymbol{\pi}$, and such a simple selection rule does not hold. 
In particular, as we show later, $S_\alpha(\boldsymbol{x}'',t|\boldsymbol{x}',0)$ can even become negative. Such solutions would imply an exponentially growing propagator and are therefore incompatible with the bounded evolution generated by Eq.~\eqref{eq:evquas}. This does not signal a breakdown of the WKB ansatz~\eqref{eq:wkb-prop}, but rather the fact that the Hamilton--Jacobi equation obtained in the limit $J\to\infty$ provides only a necessary condition for $S_\alpha$ and, in the present nonconvex setting, admits spurious branches. The additional criteria are then needed to select the physically relevant solution.

We now show that the barrier $\mathcal{A}_{i \to j}$ is determined by the instanton connecting FPs $i$ and $j$, and formulate physically motivated criteria for selecting such an instanton. As in Ref.~\cite{dykman1988quantum}, we consider the time regime where $t$ is much longer than the relaxation time within a single basin of attraction, but much shorter than the timescale of transitions between the attractors, $t \ll 1/\kappa_{i \rightarrow j}$. We further note that, although distributions $p_\alpha$ are not proper probability densities, they reproduce the magnetizations $m_k$. For a system initially in the attractor $i$, the change of $m_k$ induced by the $i \rightarrow j$ transition is thus proportional both to the propagator $K_\alpha(\boldsymbol{x}_j^*,t|\boldsymbol{x}_i^*,0)$, where $\boldsymbol{x}_{i,j}^*$ is the coordinate vector of the fixed point $i,j$, and to the transition probability $\approx \kappa_{i \rightarrow j} t$. Therefore,
$K_\alpha(\boldsymbol{x}_j^*,t|\boldsymbol{x}_i^*,0) \propto \kappa_{i \rightarrow j}t$, which by Eqs.~\eqref{eq:arrh} and~\eqref{eq:wkb-prop} implies that
\begin{align}
\mathcal{A}_{i \rightarrow j}=\lim_{t \rightarrow \infty} S_\alpha(\boldsymbol{x}_j^*,t|\boldsymbol{x}_i^*,0) \,.
\end{align}
The limit $t \rightarrow \infty$ taken after $J \rightarrow \infty$ follows from the fact that for any fixed $t$ the action becomes (quasi)stationary as $J \rightarrow \infty$~\cite{dykman1988quantum}: $\partial_t S_\alpha (\boldsymbol{x}_j^*,t|\boldsymbol{x}_i^*,0) \propto - J^{-1}\partial_t \ln K_\alpha \propto -J^{-1} \ln t \rightarrow 0$, where we used $K_\alpha \propto \kappa_{i \rightarrow j}t$. Due to Eq.~\eqref{eq:hamilton-jacobi}, it then follows that $\mathcal{H}_\alpha=0$ at the end of the trajectory $\boldsymbol{\xi}$ and thus---since Eq.~\eqref{eq:hamilton-eq} conserves the value of $\mathcal{H}_\alpha$---it is equal to $0$ along the entire trajectory $\boldsymbol{\xi}$. 
In addition, assume that the instanton $\boldsymbol{\xi}$ undergoes a continuous (but not necessarily smooth) deformation under an infinitesimal
 perturbation $\boldsymbol{x} \rightarrow \boldsymbol{x}+\delta \boldsymbol{x}$ of the initial and final points. Then, it must obey the following boundary conditions (for $t \rightarrow \infty)$:
 \begin{figure*}
    \centering
    \includegraphics[width=0.99\linewidth]{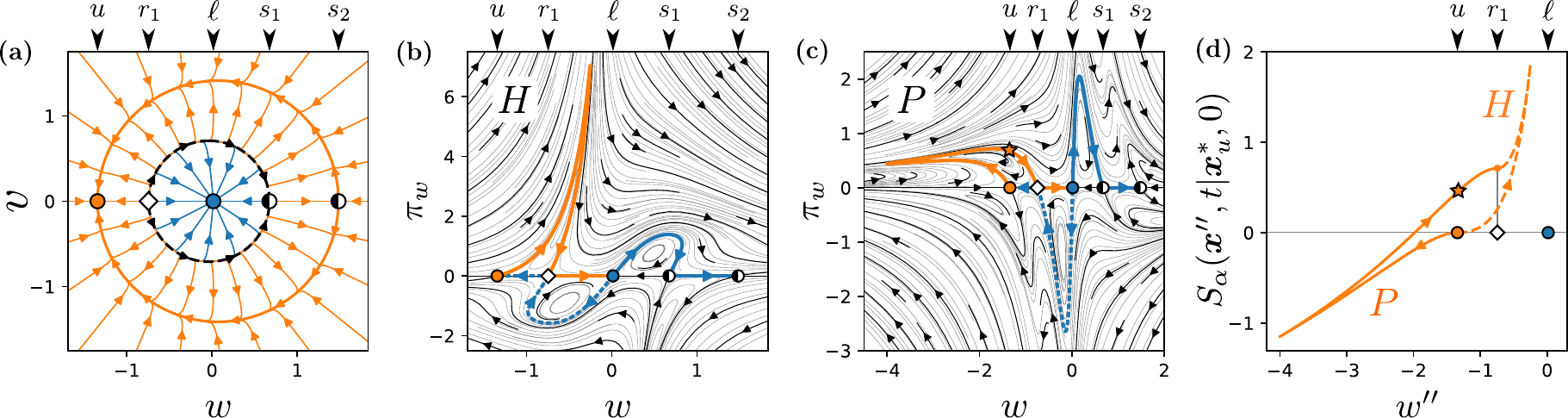}
    \caption{(a) Phase portrait of MF equations in the stereographic plane~\eqref{eq:mfster}. Arrows inside denote drift direction. Labels above denote the positions of stable FPs $\ell$, $u$ (filled circles), saddle points $s_1$, $s_2$ (half-filled circles), and source $r_1$ (empty diamond). The second source $r_2$ is outside the plot area ($w \approx 64$). (b,c) Blue (orange) lines denote the instanton trajectories  escaping the FP $\ell$ ($u$) for $H$ (b) and $P$ (c) distributions, plotted inside the velocity field of Eq.~\eqref{eq:hamilton-eq}. Arrows denote the trajectory direction. $s_2 \rightarrow u$ path not shown, as it leaves the $w$-$\pi_w$ plane. (d) The action $S_\alpha(\boldsymbol{x}'',t|\boldsymbol{x}_u^*,0)$  calculated along the $\boldsymbol{\pi} \neq \boldsymbol{0}$ instanton segment $u \rightarrow r_1$ for $\alpha=H$ (dashed line) and $\alpha=P$ (solid line). Arrows denote the trajectory direction. The star in (c,d) denotes the midpoint of the $u \to \ell$ instanton discussed in the main text. All plots for $\Omega=0.25\gamma$, $\Gamma=9\gamma$.
    }
    \label{fig:fig2}
\end{figure*}
\begin{itemize}
\item For the initial point, we note that $K_\alpha(\boldsymbol{x}_j^*,t|\boldsymbol{x}_i^*+\delta \boldsymbol{x},0) \approx K_\alpha(\boldsymbol{x}_j^*,t|\boldsymbol{x}_i^*,0)$ because such a propagator corresponds to fast relaxation $\boldsymbol{x}_i^*+\delta \boldsymbol{x} \rightarrow \boldsymbol{x}_i^*$ followed by a slow transition $\boldsymbol{x}_i^* \rightarrow \boldsymbol{x}_j^*$. Thus, by Eq.~\eqref{eq:action}, $S_\alpha(\boldsymbol{x}_j^*,t|\boldsymbol{x}_i^*+\delta \boldsymbol{x},0) -S_\alpha(\boldsymbol{x}_j^*,t|\boldsymbol{x}_i^*,0)=-\boldsymbol{\pi}(0) \cdot \delta \boldsymbol{x}=0$ to the leading order in $\delta \boldsymbol{x}$ (see Appendix~\ref{app:act-var}) and therefore $\boldsymbol{\pi}(0)=\boldsymbol{0}$.
\item Similarly, for the final point, the propagator $K_\alpha(\boldsymbol{x}'',t|\boldsymbol{x}_i^*,0)$ is locally maximal at $\boldsymbol{x}''=\boldsymbol{x}_j^*$, reflecting rapid relaxation within basin $j$.
Thus, $S_\alpha(\boldsymbol{x}_j^*+\delta \boldsymbol{x},t|\boldsymbol{x}_i^*,0) -S_\alpha(\boldsymbol{x}_j^*,t|\boldsymbol{x}_i^*,0)=\boldsymbol{\pi}(t) \cdot \delta \boldsymbol{x}=0$ and therefore $\boldsymbol{\pi}(t)=\boldsymbol{0}$.
\end{itemize}

\textit{Determining the instantons}---We now discuss how to determine the instantons satisfying the above conditions. First, note that in the $\boldsymbol{\pi}=\boldsymbol{0}$ manifold, Eq.~\eqref{eq:hamilton-eq} reduces to the MF dynamics, $d_t\boldsymbol{x}=\boldsymbol{h}(\boldsymbol{x}) \equiv \nabla_{\boldsymbol{\pi}}\mathcal{H}_\alpha(\boldsymbol{x},\boldsymbol{0})$. This follows from the expansion $L_\alpha(\boldsymbol{x},\nabla_{\boldsymbol{x}})=-\nabla_{\boldsymbol{x}} \boldsymbol{h}(\boldsymbol{x})+O(J^{-1})$~\cite{carmichael1999statistical1}, combined with Eq.~\eqref{eq:auxhamdef}. Consequently, as discussed in Appendix~\ref{app:stability-hameq}, the initial and final points, $(\boldsymbol{x},\boldsymbol{\pi})=(\boldsymbol{x}_{i,j}^*,\boldsymbol{0})$, are saddle points of the dynamical system~\eqref{eq:hamilton-eq}---stable within the $\boldsymbol{\pi}=\boldsymbol{0}$ manifold but unstable along the $\boldsymbol{\pi}\neq \boldsymbol{0}$ directions. Therefore, a trajectory $i \to j$ can be realized only by escaping the $\boldsymbol{\pi}=\boldsymbol{0}$ manifold and moving ``against'' the deterministic drift. Dynamical systems theory, previously applied to classical stochastic systems~\cite{gagrani2023action}, implies that the escape trajectory follows the unstable manifold of $(\boldsymbol{x}_i^*,\boldsymbol{0})$ to another saddle point $(\boldsymbol{x}_\nu^*,\boldsymbol{0})$ of Eq.~\eqref{eq:hamilton-eq}, forming a heteroclinic connection. Here, $\boldsymbol{x}_\nu^*$ is an unstable fixed point of the MF dynamics on the separatrix between the basins of attraction of $i$ and $j$. From there, the instanton follows the MF trajectory to the final point $(\boldsymbol{x}_j^*,\boldsymbol{0})$, which is stable in the $\boldsymbol{\pi}=\boldsymbol{0}$ manifold and thus attracts (repulses) the $\boldsymbol{\pi}=\boldsymbol{0}$ ($\boldsymbol{\pi} \neq \boldsymbol{0}$) trajectories. Because $\mathcal{H}_\alpha=0$ along the instanton trajectory, Eq.~\eqref{eq:action} reduces to $S_\alpha=\int dt \, \boldsymbol{\pi}(t)\cdot d_t\boldsymbol{x}(t)$, so that only the instanton segment with $\boldsymbol{\pi}\neq 0$ contributes to $S_\alpha$. If several such trajectories exist, the minimum action determines $\mathcal{A}_{i \rightarrow j}$.

For classical stochastic systems, the $\boldsymbol{\pi} \neq \boldsymbol{0}$ segment of the instanton can be determined using efficient minimum-action methods~\cite{grafke2019numerical,gagrani2023action,zakine2023minimum}. For nonconvex Hamiltonians, as in our case, more involved shooting~\cite{carde2026nonperturvative} or boundary-value methods~\cite{lee2024real} are usually required, as used in the companion paper~\cite{chudak2026quantum}.
However, for the present model, the analysis is simplified by its peculiar symmetry. We note that the dissipative part of the MF dynamics expressed in the stereographic coordinates [Eq.~\eqref{eq:mfster}] is rotationally symmetric around $\boldsymbol{x}=\boldsymbol{0}$, while the coherent drive $\Omega$ breaks this symmetry, inducing drift in the $+w$ direction. As a consequence, all stable and unstable FPs lie along the $v=0$ axis, where $d_t v=0$ [Fig.~\ref{fig:fig2}(a)]. Similarly, for the dynamical system~\eqref{eq:hamilton-eq}, both $d_t v$ and $d_t \pi_v$ vanish at $v,\pi_v=0$. Hence, we find that the $\boldsymbol{\pi} \neq {\boldsymbol{0}}$ part of the instanton trajectory is confined to the $w$–$\pi_w$ plane [as confirmed in Fig.~\ref{fig:fig2}(b,c)], reducing the problem to two dimensions, which enables efficient treatment via the numerical continuation method described in the Supplemental Material~\cite{supp}.

As further shown in Fig.~\ref{fig:fig2}(b,c), the $\boldsymbol{\pi} \neq \boldsymbol{0}$ parts of the instanton trajectories differ significantly for the $H$ and $P$ distributions and exhibit a folded structure, such that their projection onto the state variable $w$ may overlap with the basins of attraction of both stable FPs. The activation barrier $\mathcal{A}_{u \rightarrow \ell}$ is determined by the action along the trajectory $u \rightarrow r_1$, followed by the MF relaxation $r_1 \rightarrow \ell$. Conversely, $\mathcal{A}_{\ell \rightarrow u}$ is given by the action along $\ell \rightarrow s_1$, with subsequent relaxation along a more complex path $s_1 \rightarrow s_2 \rightarrow u$, with the last segment acquiring $v \neq 0$ [see Fig.~\ref{fig:fig2}(a)]. A more direct path $\ell \rightarrow r_1 \rightarrow u$ [blue dashed lines in Fig.~\ref{fig:fig2}(b,c)] is also allowed by Eq.~\eqref{eq:hamilton-eq}, but generates a larger action, as it opposes the drift induced by $\Omega$ (see Appendix~\ref{app:action_from_l}).

The nonconvexity of $\mathcal{H}_\alpha$, which distinguishes the quantum problem from classical stochastic systems with convex Hamiltonians~\cite{zakine2023minimum,gagrani2023action}, becomes particularly apparent in the behavior of the action $S_\alpha(\boldsymbol{x}'',t|\boldsymbol{x}_u^*,0)$ shown in Fig.~\ref{fig:fig2}(d). Unlike in classical systems, the action evaluated along the instanton trajectories is not only multivalued in $\boldsymbol{x}''$, but also nonmonotonic and can even become negative. Nevertheless, despite the different transient behavior of $S_H$ and $S_P$, for $t \rightarrow \infty$ the action converges to the same positive value $S_\alpha(\boldsymbol{x}_\ell^*,t|\boldsymbol{x}_u^*,0)$, which determines $\mathcal{A}_{u \rightarrow \ell}$. Another manifestation of nonconvexity is that the instanton need not be the minimum-action trajectory among all trajectories satisfying $\boldsymbol{x}(0)=\boldsymbol{x}_i^*$ and $\boldsymbol{x}(t)=\boldsymbol{x}_j^*$ for some $t\in(0,\infty)$. For example, a trajectory initialized at the $\pi_w>0$ midpoint of the $u\to\ell$ instanton in Fig.~\ref{fig:fig2}(c) (marked by the star) has a smaller action than the full instanton [as indicated by Fig.~\ref{fig:fig2}(d)] while still satisfying the above boundary conditions.

\textit{Comparison with QME}---To validate our approach, in Fig.~\ref{fig:mzaction}(b) we plot the obtained activation barriers. As shown, the crossing point of $\mathcal{A}_{\ell \rightarrow u}$ and $\mathcal{A}_{u \rightarrow \ell}$ at $\Gamma \approx 8.9\gamma$ coincides with the point where the finite-size results for stationary magnetization $m_{z,\text{ss}}$ deviate from the $u$ branch and begin transitioning to the $\ell$ branch. This agreement confirms that our approach correctly captures the asymptotic behavior of magnetization and the location of the first-order phase transition.

In contrast, the previously used~\cite{dutta2025quantum,xiang2025switching} semiclassical Wigner (SW) approach (which we review in~\cite{supp}) fails to reproduce this behavior. It predicts smaller activation barriers and a crossing at $\Gamma \approx 7.9\gamma$, where $m_{z,\text{ss}}$ remains localized around the $u$ branch. This discrepancy arises because the SW approach truncates third- and higher-order derivative terms in the corresponding $L_\alpha$ operator (describing non-Gaussian fluctuations), which---according to Eq.~\eqref{eq:auxhamdef}---contribute to $\mathcal{H}_\alpha$. The breakdown of such truncations in evaluating activation barriers has also been reported for classical stochastic~\cite{hanggi1988bistability,gaveau1997master,kessler2007extinction} and quantum bosonic systems~\cite{drummond1989quantum,kinsler1991quantum}.

\begin{figure}
    \centering
    \includegraphics[width=0.9\linewidth]{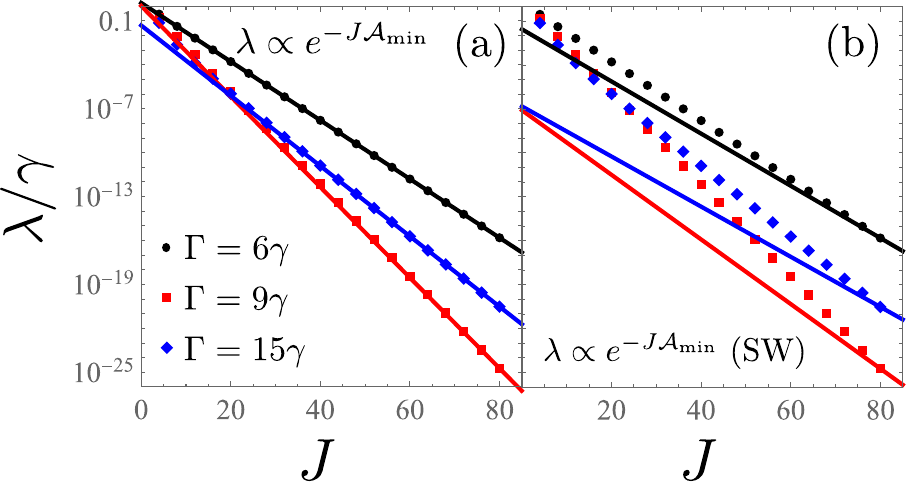}
    \caption{(a) Dots: Liouvillian gap $\lambda$ calculated using QME for different $\Gamma$, plotted in the logarithmic scale; calculation details in~\cite{supp}. Lines: the predicted scaling $\lambda \propto \exp(-J \mathcal{A}_\text{min})$ set to cross the point for $J=80$. (b) The same plot, but with $\mathcal{A}_\text{min}$ given by the SW approach. Parameter $\Omega=0.25 \gamma$.
    }
    \label{fig:liovgap}
\end{figure}
\textit{Liouvillian gap}---For further validation, we consider the slowest relaxation timescale, characterized by the Liouvillian gap $\lambda$. It is determined by ordering the eigenvalues $\lambda_i$ of $\mathcal{L} \hat{\rho}_i=\lambda_i \hat{\rho}_i$ as $0=- \re \lambda_0 \leq -\re \lambda_1\leq \ldots$. The Liouvillian gap is then defined as $\lambda \equiv -\re \lambda_1$~\cite{minganti2018spectral}. In bistable systems, it sets the inverse timescale for transitions from the less probable to the more probable attractor, such that $\lambda \approx \max(\kappa_{\ell \rightarrow u},\kappa_{u \rightarrow \ell}) \asymp \exp(-J \mathcal{A}_\text{min}) $, with $\mathcal{A}_\text{min} \equiv \min(\mathcal{A}_{\ell \rightarrow u},\mathcal{A}_{u \rightarrow \ell})$~\cite{macieszczak2021theory,xiang2025switching,carde2026nonperturvative}. As shown in Fig.~\ref{fig:liovgap}, for large $J$, $\lambda$ indeed decays exponentially with $J$. Its decay rate is well captured by our method but is significantly underestimated by the SW approach.

Using $\lambda \asymp \exp(-J \mathcal{A}_\text{min})$ we also define the estimator
\begin{align} \label{eq:aminest}
\tilde{\mathcal{A}}_{\text{min},J} \equiv \frac{1}{4} \ln \frac{\lambda(J-4)}{\lambda(J)} \approx  \mathcal{A}_{\text{min}} \,.
\end{align}
As shown in Fig.~\ref{fig:mzaction}(b), this estimator agrees well with our prediction of $\mathcal{A}_\text{min}$, but not with that of the SW approach.

\textit{Validity for larger $\Omega$}---Finally, to further confirm our conclusions, the Supplemental Material~\cite{supp} shows that they also hold for a larger value of $\Omega = 0.5\gamma$.


\textit{Concluding remarks}---Our work demonstrates that an instanton approach based on exact quantum quasiprobability dynamics, beyond the semiclassical approximation, can accurately characterize metastability in systems governed by quantum master equations (QMEs). The companion paper~\cite{chudak2026quantum} further supports this conclusion through applications to microscopically motivated models with collective and local dissipation. Although we focused on collective spin systems, our theoretical framework applies more broadly to systems admitting quasiprobability dynamics of the form in Eq.~\eqref{eq:evquas}. These include, e.g., spin-boson complexes~\cite{debecker2024controlling}, whose quasiprobability representations have been considered in Refs.~\cite{haken1967quantum,carhmichael1986quantum,mandt2015stochastic,carmichael2008statistical2}. The approach also naturally extends to feedback-controlled systems, where the QME is coupled to classical Langevin dynamics for the measurement record~\cite{andersson2022quantum}. In a quasiprobability representation, this yields a unified partial differential equation for the joint state of the system and feedback device, allowing Eq.~\eqref{eq:evquas} to be applied directly.

\begin{acknowledgments}
\textit{Acknowledgments}---K.P.\ and M.C.\ acknowledge the financial support of the National Science Centre, Poland, under the project No.\ 2023/51/D/ST3/01203, and M.E.\ of the Fond National de la Recherche-FNR, Luxembourg, CORE project NEQPHASETRANS (C24/MS/18933049).
\end{acknowledgments}

\textit{Data availability}---The plotted data, numerical calculations, and derivations of $\mathcal{H}_\alpha$ that support the findings of this article are openly
available at~\cite{zenodo,github}.


\appendix

\begin{center}
  \large \bf End Matter
\end{center}

\section{MF dynamics} \label{app:mf}
Here, we sketch the derivation of the MF equations~\eqref{eq:meanfield}, which is rigorously proven in Refs.~\cite{alicki1983nonlinear,benatti2016non,benatti2018quantum,fiorelli2023mean}. The starting point is the Heisenberg evolution of spin operators
\begin{align} \label{eq:heis}
d_t \hat{J}_k=\mathcal{L}^\dagger \hat{J}_k \,,
\end{align}
where $\mathcal{L}^\dagger$ is the superoperator adjoint to $\mathcal{L}$. It is obtained from $\mathcal{L}$ by replacement $\hat{H} \equiv \Omega \hat{J}_x \rightarrow -\hat{H}$ and $\mathcal{D}[\hat{A}] \bullet \rightarrow \mathcal{D}^\dagger[\hat{A}] \bullet\equiv\hat{A}^\dagger \bullet \hat{A}-\{\hat{A}^\dagger \hat{A},\bullet\}/2$. This leads to the equations of motion~\cite{dutta2025quantum}
\begin{align} \nonumber
d_t \hat{J}_+&=- i\Omega \hat{J}_z-\frac{\gamma}{J} \hat{J}_z \hat{J}_++\frac{\Gamma}{J^3} \hat{J}_+ \left(\hat{J}_z^3+2\hat{J}_z^2-\hat{J}_-\hat{J}_+/2 \right) \,, \\ \nonumber
d_t \hat{J}_-&=i\Omega \hat{J}_z-\frac{\gamma}{J} \hat{J}_- \hat{J}_z+\frac{\Gamma}{J^3}  \left(\hat{J}_z^3+2\hat{J}_z^2-\hat{J}_-\hat{J}_+/2 \right)\hat{J}_- \,, \\ \label{eq:mfder}
d_t \hat{J}_z&=-\frac{i \Omega}{2}(\hat{J}_+-\hat{J}_-)+\frac{\gamma}{J} \hat{J}_- \hat{J}_+-\frac{\Gamma}{J^3} \hat{J}_+ \hat{J}_- \hat{J}_z^2 \,.
\end{align}
The MF dynamics~\eqref{eq:meanfield} is then obtained by replacing operators with classical variables $\hat{J}_\pm \rightarrow J(m_x \pm i m_y)$ and $\hat{J}_z \rightarrow J m_z$, dividing both sides of Eq.~\eqref{eq:mfder} by $J$, and omitting terms of order $O(1/J)$.

\section{Stability analysis} \label{app:stability}
To determine whether the FPs $d_t m_k=0$ of the MF dynamics~\eqref{eq:meanfield} are stable or unstable, we rewrite the MF equations in stereographic coordinates:
\begin{align} \nonumber
d_t v&=\Omega vw+\gamma v-\Gamma v \frac{(v^2+w^2-1)^2}{(v^2+w^2+1)^2} \,, \\ \label{eq:mfster}
d_t w&=\frac{\Omega}{2} (1-v^2+w^2)+\gamma w-\Gamma w \frac{(v^2+w^2-1)^2}{(v^2+w^2+1)^2} \,,
\end{align}
with $v=m_x/(1-m_z)$, $w=m_y/(1-m_z)$.
The stability of FP $d_t \boldsymbol{x}=\boldsymbol{0} \vert_{\boldsymbol{x}=\boldsymbol{x}^*}$ is determined by the Jacobian
\begin{align} \label{eq:mfjacobian}
\mathbb{J} = \begin{pmatrix} \partial_v \dot{v} & \partial_v \dot{w} \\ \partial_w \dot{v} & \partial_w \dot{w} \end{pmatrix}_{\boldsymbol{x}=\boldsymbol{x}^*} \,.
\end{align}
The FP is stable if both of its eigenvalues have negative real parts, while it is a saddle (source) when one (both) of them is (are) positive.

\section{Derivation of $L_\alpha$ and $\mathcal{H}_\alpha$} \label{app:derevquas}
Here we review the method to derive differential operators $L_\alpha(\boldsymbol{x},\nabla_{\boldsymbol{x}})$, and thus auxiliary Hamiltonians $\mathcal{H}_\alpha(\boldsymbol{x},\boldsymbol{\pi})$. To that end, the action of the spin operators on $|\boldsymbol{x}\rangle \langle \boldsymbol{x}|$, where $|\boldsymbol{x}\rangle$ is the coherent spin state expressed in the stereographic coordinates, is represented by the corresponding differential operators~\cite{narducci1975multitime,altland2012quantum,mandt2015stochastic,burkle2020probabilistic}:
\begin{align}
\hat{A} |\boldsymbol{x}\rangle \langle \boldsymbol{x}| \equiv \mathcal{J}[\hat{A}]|\boldsymbol{x}\rangle \langle \boldsymbol{x}| \,, \quad |\boldsymbol{x}\rangle \langle \boldsymbol{x}| \hat{A} \equiv \mathcal{J}^*[\hat{A}^\dagger]|\boldsymbol{x}\rangle \langle \boldsymbol{x}| \,,
\end{align}
where
\begin{align} \nonumber
\mathcal{J}[\hat{J}_+]&=\frac{1}{2} \left[ \frac{\partial}{\partial v}+i\frac{\partial}{\partial w}+\frac{4J(v+iw)}{v^2+w^2+1} \right] \,, \\ \nonumber
\mathcal{J}[\hat{J}_-]&=\frac{1}{2} \left[ -(v-iw)^2\left(\frac{\partial}{\partial v}+i\frac{\partial}{\partial w}\right)+\frac{4J(v-iw)}{v^2+w^2+1} \right] \,, \\ \label{eq:corrules}
\mathcal{J}[\hat{J}_z]&=\frac{1}{2} \left[(v-iw)\left(\frac{\partial}{\partial v}+i\frac{\partial}{\partial w}\right)+\frac{2J(v^2+w^2-1)}{v^2+w^2+1} \right] \,.
\end{align}
Here we adapted operators from Ref.~\cite{burkle2020probabilistic} transformed into stereographic coordinates. The operators $L_\alpha(\boldsymbol{x},\nabla_{\boldsymbol{x}})$ are given by the correspondence rules~\cite{burkle2020probabilistic,altland2012quantum,mandt2015stochastic}
\begin{subequations}
\begin{align}
L_H(\boldsymbol{x},\nabla_{\boldsymbol{x}}) |\boldsymbol{x}\rangle \langle \boldsymbol{x}| &\equiv \mathcal{L}^\dagger |\boldsymbol{x}\rangle \langle \boldsymbol{x}| \,, \\
L_P^\dagger(\boldsymbol{x},\nabla_{\boldsymbol{x}}) |\boldsymbol{x}\rangle \langle \boldsymbol{x}| &\equiv \mathcal{L}|\boldsymbol{x}\rangle \langle \boldsymbol{x}| \,,
\end{align}
\end{subequations}
with $\mathcal{L}^\dagger$ defined below Eq.~\eqref{eq:heis} and
where $L_P^\dagger$ is the adjoint of $L_P$. Expanding $L_P^\dagger$ as
\begin{align}
L_P^\dagger(\boldsymbol{x},\nabla_{\boldsymbol{x}})=\sum_{a,b} L_P^{(a,b)} (\boldsymbol{x}) \frac{\partial^a}{\partial v^a} \frac{\partial^b}{\partial w^b} \,,
\end{align}
the operator $L_P$ can be obtained as
\begin{align}
L_P(\boldsymbol{x},\nabla_{\boldsymbol{x}})=\sum_{a,b} (-1)^{a+b}\frac{\partial^a}{\partial v^a} \frac{\partial^b}{\partial w^b} L_P^{(a,b)} (\boldsymbol{x}) \,.
\end{align}
Employing Eq.~\eqref{eq:auxhamdef}, the Hamiltonian $\mathcal{H}_P$ can also be obtained directly from $L_P^\dagger$ as $\mathcal{H}_P(\boldsymbol{x},\boldsymbol{\pi})=\lim_{J \rightarrow \infty} J^{-1} L_P^\dagger(\boldsymbol{x},J \boldsymbol{\pi})$. In general, to derive $\mathcal{H}_\alpha$, it is most easy to perform replacement $\partial_{v,w} \rightarrow - J \pi_{v,w}$ ($\alpha=H$) or $\partial_{v,w} \rightarrow J \pi_{v,w}$ ($\alpha=P$) already at the level of operators~\eqref{eq:corrules}, so one does not have to take care with operator ordering. Since the explicit expressions for $\mathcal{H}_\alpha$ are very extensive, instead of copying them, we make the Wolfram Mathematica notebooks used to derive them available at~\cite{zenodo}; they can be read, e.g., using free-of-charge Wolfram Player. We also copy their prints in~\cite{supp}.

\section{Action variation} \label{app:act-var}
Variations of the action with respect to the endpoints are determined by the conjugate momenta,
\begin{align}
S_\alpha(\boldsymbol{x}'',t|\boldsymbol{x}'+\delta \boldsymbol{x},0)
-
S_\alpha(\boldsymbol{x}'',t|\boldsymbol{x}',0)
=
- \boldsymbol{\pi}(0)\cdot \delta \boldsymbol{x},
\end{align}
and analogously for the final point. To show this, we consider a generic perturbation of the trajectory
\begin{align}
\boldsymbol{x}(t') \to \boldsymbol{x}(t')+\delta \boldsymbol{x}(t'), 
\qquad 
\boldsymbol{\pi}(t') \to \boldsymbol{\pi}(t')+\delta \boldsymbol{\pi}(t').
\end{align}
The variation of the action [see Eq.~\eqref{eq:action}] reads
\begin{align} 
&\delta S_\alpha= \\ \nonumber  & \int_0^t dt' \big[
\boldsymbol{\pi}\cdot \delta \dot{\boldsymbol{x}}
+ \dot{\boldsymbol{x}}\cdot \delta \boldsymbol{\pi}
-(\nabla_{\boldsymbol{x}} \mathcal{H}_\alpha)\cdot \delta\boldsymbol{x}
-(\nabla_{\boldsymbol{\pi}} \mathcal{H}_\alpha)\cdot \delta\boldsymbol{\pi}
\big].
\end{align}
Integrating $\boldsymbol{\pi}\cdot \delta \dot{\boldsymbol{x}}$ by parts, we obtain
\begin{align} 
\delta S_\alpha &= \boldsymbol{\pi}(t)\cdot \delta \boldsymbol{x}(t)
- \boldsymbol{\pi}(0)\cdot \delta \boldsymbol{x}(0) \\
&+ \int_0^t dt' \big[
(\dot{\boldsymbol{x}}-\nabla_{\boldsymbol{\pi}} \mathcal{H}_\alpha)\cdot \delta \boldsymbol{\pi}
-(\dot{\boldsymbol{\pi}}+\nabla_{\boldsymbol{x}} \mathcal{H}_\alpha)\cdot \delta\boldsymbol{x}
\big].\nonumber
\end{align}
The integral vanishes due to Eq.~\eqref{eq:hamilton-eq}, so that
\begin{align} 
\delta S_\alpha = \boldsymbol{\pi}(t)\cdot \delta \boldsymbol{x}(t)
- \boldsymbol{\pi}(0)\cdot \delta \boldsymbol{x}(0).
\end{align}
For variations of the final point at fixed initial condition ($\delta \boldsymbol{x}(0)=0$), this implies
\begin{align}\label{GradSasPi}
\nabla_{\boldsymbol{x}''}S_\alpha(\boldsymbol{x}'',t|\boldsymbol{x}',0) = \boldsymbol{\pi}(t),
\end{align}
and similarly, for variations of the initial point at fixed final condition, one obtains
\begin{align}
\nabla_{\boldsymbol{x}'}S_\alpha(\boldsymbol{x}'',t|\boldsymbol{x}',0) = -\boldsymbol{\pi}(0).
\end{align}

\section{Stability analysis of Eq.~\eqref{eq:hamilton-eq}} \label{app:stability-hameq}
Here we analyze the stability of the FPs of Eq.~\eqref{eq:hamilton-eq}. We define the vector $\boldsymbol{q}=(\boldsymbol{x},\boldsymbol{\pi})$. Each MF fixed point $\boldsymbol{x}^*$ then corresponds to a FP $\boldsymbol{q}^*=(\boldsymbol{x}^*,\boldsymbol{0})$ of Eq.~\eqref{eq:hamilton-eq}. The dynamics of a small displacement $\delta \boldsymbol{q}\equiv \boldsymbol{q}-\boldsymbol{q}^*$ is governed by the linearized equation
\begin{align}
\delta \dot{\boldsymbol{q}}=\delta \boldsymbol{q} \cdot \mathbb{K} \,,
\end{align}
where
\begin{align}
\mathbb{K} \equiv [\partial_{q_k} \dot{q}_l]_{k,l} \, \big\vert_{\boldsymbol{q}=\boldsymbol{q}^*}
=\begin{pmatrix} \mathbb{J} & 0 \\ 2\mathbb{D} & -\mathbb{J}^\intercal \end{pmatrix} \,,
\end{align}
is the corresponding Jacobian matrix. Here, $\mathbb{J}$ is the MF Jacobian defined in Eq.~\eqref{eq:mfjacobian}, while 
\begin{align}
\mathbb{D} \equiv \frac{1}{2} \begin{pmatrix} \partial_{\pi_v}^2 \mathcal{H}_\alpha & \partial_{\pi_v} \partial_{\pi_w} \mathcal{H}_\alpha \\ \partial_{\pi_v} \partial_{\pi_w} \mathcal{H}_\alpha & \partial_{\pi_w}^2 \mathcal{H}_\alpha \end{pmatrix}_{\boldsymbol{q}=\boldsymbol{q}^*}
\end{align}
is the diffusion matrix. In deriving this, we used $\dot{\boldsymbol{x}}=\nabla_{\boldsymbol{\pi}}\mathcal{H}_\alpha$ [Eq.~\eqref{eq:hamilton-eq}]. 

The matrix $\mathbb{K}$ has eigenvalues $\{a_1,a_2,-a_1,-a_2\}$, so that each FP $\boldsymbol{q}^*$ is a saddle with an equal number of stable and unstable directions. Here, $a_{1,2}$ are the eigenvalues of $\mathbb{J}$. The associated left eigenvectors take the form $(\boldsymbol{j}_{1,2},\boldsymbol{0})$, where $\boldsymbol{j}_{1,2}$ are the left eigenvectors of $\mathbb{J}$; consequently, they have zero $\boldsymbol{\pi}$ component. In contrast, the remaining eigenvectors have a nonzero $\boldsymbol{\pi}$ component. 

\begin{figure}
    \centering
    \includegraphics[width=0.99\linewidth]{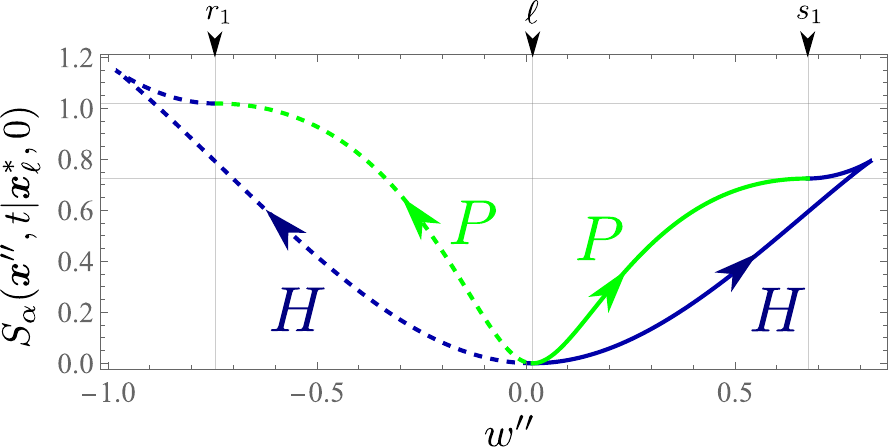}
    \caption{The action $S_\alpha(\boldsymbol{x}'',t|\boldsymbol{x}_\ell^*,0)$  calculated along the $\boldsymbol{\pi} \neq \boldsymbol{0}$ instanton segments $\ell \rightarrow r_1$ (dashed lines) and $\ell \rightarrow s_1$ (solid lines) for $\alpha=H$ (dark blue) and $\alpha=P$ (green). Arrows denote the trajectory directions. Parameters as in Fig.~\ref{fig:fig2}: $\Omega=0.25\gamma$, $\Gamma=9\gamma$.
    }
    \label{fig:action_from_l}
\end{figure}%
For saddles corresponding to stable MF fixed points, we have $\re \, a_{1,2}<0$, which implies that the $\boldsymbol{\pi}=\boldsymbol{0}$ directions are stable, while the $\boldsymbol{\pi}\neq\boldsymbol{0}$ directions are unstable. As a result, an instanton trajectory escaping such a point must have a nonzero $\boldsymbol{\pi}$ component. In contrast, such a point can be approached only along an instanton segment with $\boldsymbol{\pi}=\boldsymbol{0}$, corresponding to the MF relaxation path. As follows from dynamical systems theory, the trajectory linking two saddles is composed of segments corresponding to heteroclinic connections between saddle points~\cite{gagrani2023action}. Therefore, the $\boldsymbol{\pi}\neq\boldsymbol{0}$ and $\boldsymbol{\pi}=\boldsymbol{0}$ instanton segments connect in a saddle corresponding to an unstable MF fixed point, which possesses at least one stable (unstable) direction with $\boldsymbol{\pi}\neq\boldsymbol{0}$ ($\boldsymbol{\pi}=\boldsymbol{0}$).

\section{Action for escaping $\ell$} \label{app:action_from_l}
In Fig.~\ref{fig:action_from_l} we plot the evolution of action $S_\alpha$ along the instanton segments $\ell \to r_1$ and $\ell \to s_1$, demonstrating that the former path is associated with larger action.

\bibliography{bibliography}

\begin{thebibliography}{70}%
\makeatletter
\providecommand \@ifxundefined [1]{%
 \@ifx{#1\undefined}
}%
\providecommand \@ifnum [1]{%
 \ifnum #1\expandafter \@firstoftwo
 \else \expandafter \@secondoftwo
 \fi
}%
\providecommand \@ifx [1]{%
 \ifx #1\expandafter \@firstoftwo
 \else \expandafter \@secondoftwo
 \fi
}%
\providecommand \natexlab [1]{#1}%
\providecommand \enquote  [1]{``#1''}%
\providecommand \bibnamefont  [1]{#1}%
\providecommand \bibfnamefont [1]{#1}%
\providecommand \citenamefont [1]{#1}%
\providecommand \href@noop [0]{\@secondoftwo}%
\providecommand \href [0]{\begingroup \@sanitize@url \@href}%
\providecommand \@href[1]{\@@startlink{#1}\@@href}%
\providecommand \@@href[1]{\endgroup#1\@@endlink}%
\providecommand \@sanitize@url [0]{\catcode `\\12\catcode `\$12\catcode `\&12\catcode `\#12\catcode `\^12\catcode `\_12\catcode `\%12\relax}%
\providecommand \@@startlink[1]{}%
\providecommand \@@endlink[0]{}%
\providecommand \url  [0]{\begingroup\@sanitize@url \@url }%
\providecommand \@url [1]{\endgroup\@href {#1}{\urlprefix }}%
\providecommand \urlprefix  [0]{URL }%
\providecommand \Eprint [0]{\href }%
\providecommand \doibase [0]{https://doi.org/}%
\providecommand \selectlanguage [0]{\@gobble}%
\providecommand \bibinfo  [0]{\@secondoftwo}%
\providecommand \bibfield  [0]{\@secondoftwo}%
\providecommand \translation [1]{[#1]}%
\providecommand \BibitemOpen [0]{}%
\providecommand \bibitemStop [0]{}%
\providecommand \bibitemNoStop [0]{.\EOS\space}%
\providecommand \EOS [0]{\spacefactor3000\relax}%
\providecommand \BibitemShut  [1]{\csname bibitem#1\endcsname}%
\let\auto@bib@innerbib\@empty
\bibitem [{\citenamefont {Macieszczak}\ \emph {et~al.}(2021)\citenamefont {Macieszczak}, \citenamefont {Rose}, \citenamefont {Lesanovsky},\ and\ \citenamefont {Garrahan}}]{macieszczak2021theory}%
  \BibitemOpen
  \bibfield  {author} {\bibinfo {author} {\bibfnamefont {K.}~\bibnamefont {Macieszczak}}, \bibinfo {author} {\bibfnamefont {D.~C.}\ \bibnamefont {Rose}}, \bibinfo {author} {\bibfnamefont {I.}~\bibnamefont {Lesanovsky}},\ and\ \bibinfo {author} {\bibfnamefont {J.~P.}\ \bibnamefont {Garrahan}},\ }\bibfield  {title} {\bibinfo {title} {Theory of classical metastability in open quantum systems},\ }\href {https://doi.org/10.1103/PhysRevResearch.3.033047} {\bibfield  {journal} {\bibinfo  {journal} {Phys. Rev. Research}\ }\textbf {\bibinfo {volume} {3}},\ \bibinfo {pages} {033047} (\bibinfo {year} {2021})}\BibitemShut {NoStop}%
\bibitem [{\citenamefont {Carr}\ \emph {et~al.}(2013)\citenamefont {Carr}, \citenamefont {Ritter}, \citenamefont {Wade}, \citenamefont {Adams},\ and\ \citenamefont {Weatherill}}]{carr2013nonequilibrium}%
  \BibitemOpen
  \bibfield  {author} {\bibinfo {author} {\bibfnamefont {C.}~\bibnamefont {Carr}}, \bibinfo {author} {\bibfnamefont {R.}~\bibnamefont {Ritter}}, \bibinfo {author} {\bibfnamefont {C.}~\bibnamefont {Wade}}, \bibinfo {author} {\bibfnamefont {C.~S.}\ \bibnamefont {Adams}},\ and\ \bibinfo {author} {\bibfnamefont {K.~J.}\ \bibnamefont {Weatherill}},\ }\bibfield  {title} {\bibinfo {title} {{Nonequilibrium Phase Transition in a Dilute Rydberg Ensemble}},\ }\href {https://doi.org/10.1103/PhysRevLett.111.113901} {\bibfield  {journal} {\bibinfo  {journal} {Phys. Rev. Lett.}\ }\textbf {\bibinfo {volume} {111}},\ \bibinfo {pages} {113901} (\bibinfo {year} {2013})}\BibitemShut {NoStop}%
\bibitem [{\citenamefont {Rodriguez}\ \emph {et~al.}(2017)\citenamefont {Rodriguez}, \citenamefont {Casteels}, \citenamefont {Storme} \emph {et~al.}}]{rodriguez2017probing}%
  \BibitemOpen
  \bibfield  {author} {\bibinfo {author} {\bibfnamefont {S.}~\bibnamefont {Rodriguez}}, \bibinfo {author} {\bibfnamefont {W.}~\bibnamefont {Casteels}}, \bibinfo {author} {\bibfnamefont {F.}~\bibnamefont {Storme}}, \emph {et~al.},\ }\bibfield  {title} {\bibinfo {title} {Probing a dissipative phase transition via dynamical optical hysteresis},\ }\href {https://doi.org/10.1103/PhysRevLett.118.247402} {\bibfield  {journal} {\bibinfo  {journal} {Phys. Rev. Lett.}\ }\textbf {\bibinfo {volume} {118}},\ \bibinfo {pages} {247402} (\bibinfo {year} {2017})}\BibitemShut {NoStop}%
\bibitem [{\citenamefont {Fink}\ \emph {et~al.}(2018)\citenamefont {Fink}, \citenamefont {Schade}, \citenamefont {H{\"o}fling}, \citenamefont {Schneider},\ and\ \citenamefont {Imamoglu}}]{fink2018signatures}%
  \BibitemOpen
  \bibfield  {author} {\bibinfo {author} {\bibfnamefont {T.}~\bibnamefont {Fink}}, \bibinfo {author} {\bibfnamefont {A.}~\bibnamefont {Schade}}, \bibinfo {author} {\bibfnamefont {S.}~\bibnamefont {H{\"o}fling}}, \bibinfo {author} {\bibfnamefont {C.}~\bibnamefont {Schneider}},\ and\ \bibinfo {author} {\bibfnamefont {A.}~\bibnamefont {Imamoglu}},\ }\bibfield  {title} {\bibinfo {title} {Signatures of a dissipative phase transition in photon correlation measurements},\ }\href {https://doi.org/10.1038/s41567-017-0020-9} {\bibfield  {journal} {\bibinfo  {journal} {Nat. Phys.}\ }\textbf {\bibinfo {volume} {14}},\ \bibinfo {pages} {365} (\bibinfo {year} {2018})}\BibitemShut {NoStop}%
\bibitem [{\citenamefont {Chen}\ \emph {et~al.}(2023)\citenamefont {Chen}, \citenamefont {Fischer}, \citenamefont {Nojiri}, \citenamefont {Renger}, \citenamefont {Xie}, \citenamefont {Partanen}, \citenamefont {Pogorzalek}, \citenamefont {Fedorov}, \citenamefont {Marx}, \citenamefont {Deppe} \emph {et~al.}}]{chen2023quantum}%
  \BibitemOpen
  \bibfield  {author} {\bibinfo {author} {\bibfnamefont {Q.-M.}\ \bibnamefont {Chen}}, \bibinfo {author} {\bibfnamefont {M.}~\bibnamefont {Fischer}}, \bibinfo {author} {\bibfnamefont {Y.}~\bibnamefont {Nojiri}}, \bibinfo {author} {\bibfnamefont {M.}~\bibnamefont {Renger}}, \bibinfo {author} {\bibfnamefont {E.}~\bibnamefont {Xie}}, \bibinfo {author} {\bibfnamefont {M.}~\bibnamefont {Partanen}}, \bibinfo {author} {\bibfnamefont {S.}~\bibnamefont {Pogorzalek}}, \bibinfo {author} {\bibfnamefont {K.~G.}\ \bibnamefont {Fedorov}}, \bibinfo {author} {\bibfnamefont {A.}~\bibnamefont {Marx}}, \bibinfo {author} {\bibfnamefont {F.}~\bibnamefont {Deppe}}, \emph {et~al.},\ }\bibfield  {title} {\bibinfo {title} {Quantum behavior of the {D}uffing oscillator at the dissipative phase transition},\ }\href {https://doi.org/10.1038/s41467-023-38217-x} {\bibfield  {journal} {\bibinfo  {journal} {Nat. Commun.}\ }\textbf {\bibinfo {volume} {14}},\ \bibinfo {pages} {2896} (\bibinfo {year} {2023})}\BibitemShut {NoStop}%
\bibitem [{\citenamefont {Beaulieu}\ \emph {et~al.}(2025)\citenamefont {Beaulieu}, \citenamefont {Minganti}, \citenamefont {Frasca}, \citenamefont {Savona}, \citenamefont {Felicetti}, \citenamefont {Di~Candia},\ and\ \citenamefont {Scarlino}}]{beaulieu2025observation}%
  \BibitemOpen
  \bibfield  {author} {\bibinfo {author} {\bibfnamefont {G.}~\bibnamefont {Beaulieu}}, \bibinfo {author} {\bibfnamefont {F.}~\bibnamefont {Minganti}}, \bibinfo {author} {\bibfnamefont {S.}~\bibnamefont {Frasca}}, \bibinfo {author} {\bibfnamefont {V.}~\bibnamefont {Savona}}, \bibinfo {author} {\bibfnamefont {S.}~\bibnamefont {Felicetti}}, \bibinfo {author} {\bibfnamefont {R.}~\bibnamefont {Di~Candia}},\ and\ \bibinfo {author} {\bibfnamefont {P.}~\bibnamefont {Scarlino}},\ }\bibfield  {title} {\bibinfo {title} {Observation of first-and second-order dissipative phase transitions in a two-photon driven {K}err resonator},\ }\href {https://doi.org/10.1038/s41467-025-56830-w} {\bibfield  {journal} {\bibinfo  {journal} {Nat. Commun.}\ }\textbf {\bibinfo {volume} {16}},\ \bibinfo {pages} {1954} (\bibinfo {year} {2025})}\BibitemShut {NoStop}%
\bibitem [{\citenamefont {Noachtar}\ and\ \citenamefont {Clerk}(2026)}]{noachtar2026exact}%
  \BibitemOpen
  \bibfield  {author} {\bibinfo {author} {\bibfnamefont {D.~D.}\ \bibnamefont {Noachtar}}\ and\ \bibinfo {author} {\bibfnamefont {A.~A.}\ \bibnamefont {Clerk}},\ }\href@noop {} {\bibinfo {title} {Exact metastability in a class of driven-dissipative quantum many-body systems}} (\bibinfo {year} {2026}),\ \Eprint {https://arxiv.org/abs/2606.07736} {arXiv:2606.07736 [quant-ph]} \BibitemShut {NoStop}%
\bibitem [{\citenamefont {Drummond}\ and\ \citenamefont {Kinsler}(1989)}]{drummond1989quantum}%
  \BibitemOpen
  \bibfield  {author} {\bibinfo {author} {\bibfnamefont {P.~D.}\ \bibnamefont {Drummond}}\ and\ \bibinfo {author} {\bibfnamefont {P.}~\bibnamefont {Kinsler}},\ }\bibfield  {title} {\bibinfo {title} {Quantum tunneling and thermal activation in the parametric oscillator},\ }\href {https://doi.org/10.1103/PhysRevA.40.4813} {\bibfield  {journal} {\bibinfo  {journal} {Phys. Rev. A}\ }\textbf {\bibinfo {volume} {40}},\ \bibinfo {pages} {4813} (\bibinfo {year} {1989})}\BibitemShut {NoStop}%
\bibitem [{\citenamefont {Kinsler}\ and\ \citenamefont {Drummond}(1991)}]{kinsler1991quantum}%
  \BibitemOpen
  \bibfield  {author} {\bibinfo {author} {\bibfnamefont {P.}~\bibnamefont {Kinsler}}\ and\ \bibinfo {author} {\bibfnamefont {P.~D.}\ \bibnamefont {Drummond}},\ }\bibfield  {title} {\bibinfo {title} {{Quantum dynamics of the parametric oscillator}},\ }\href {https://doi.org/10.1103/PhysRevA.43.6194} {\bibfield  {journal} {\bibinfo  {journal} {Phys. Rev. A}\ }\textbf {\bibinfo {volume} {43}},\ \bibinfo {pages} {6194} (\bibinfo {year} {1991})}\BibitemShut {NoStop}%
\bibitem [{\citenamefont {Lee}\ \emph {et~al.}(2012)\citenamefont {Lee}, \citenamefont {H\"affner},\ and\ \citenamefont {Cross}}]{lee2012colllective}%
  \BibitemOpen
  \bibfield  {author} {\bibinfo {author} {\bibfnamefont {T.~E.}\ \bibnamefont {Lee}}, \bibinfo {author} {\bibfnamefont {H.}~\bibnamefont {H\"affner}},\ and\ \bibinfo {author} {\bibfnamefont {M.~C.}\ \bibnamefont {Cross}},\ }\bibfield  {title} {\bibinfo {title} {Collective quantum jumps of {R}ydberg atoms},\ }\href {https://doi.org/10.1103/PhysRevLett.108.023602} {\bibfield  {journal} {\bibinfo  {journal} {Phys. Rev. Lett.}\ }\textbf {\bibinfo {volume} {108}},\ \bibinfo {pages} {023602} (\bibinfo {year} {2012})}\BibitemShut {NoStop}%
\bibitem [{\citenamefont {Minganti}\ \emph {et~al.}(2018)\citenamefont {Minganti}, \citenamefont {Biella}, \citenamefont {Bartolo},\ and\ \citenamefont {Ciuti}}]{minganti2018spectral}%
  \BibitemOpen
  \bibfield  {author} {\bibinfo {author} {\bibfnamefont {F.}~\bibnamefont {Minganti}}, \bibinfo {author} {\bibfnamefont {A.}~\bibnamefont {Biella}}, \bibinfo {author} {\bibfnamefont {N.}~\bibnamefont {Bartolo}},\ and\ \bibinfo {author} {\bibfnamefont {C.}~\bibnamefont {Ciuti}},\ }\bibfield  {title} {\bibinfo {title} {Spectral theory of {L}iouvillians for dissipative phase transitions},\ }\href {https://doi.org/10.1103/PhysRevA.98.042118} {\bibfield  {journal} {\bibinfo  {journal} {Phys. Rev. A}\ }\textbf {\bibinfo {volume} {98}},\ \bibinfo {pages} {042118} (\bibinfo {year} {2018})}\BibitemShut {NoStop}%
\bibitem [{\citenamefont {Roberts}\ and\ \citenamefont {Clerk}(2023)}]{roberts2023exact}%
  \BibitemOpen
  \bibfield  {author} {\bibinfo {author} {\bibfnamefont {D.}~\bibnamefont {Roberts}}\ and\ \bibinfo {author} {\bibfnamefont {A.~A.}\ \bibnamefont {Clerk}},\ }\bibfield  {title} {\bibinfo {title} {Exact solution of the infinite-range dissipative transverse-field {I}sing model},\ }\href {https://doi.org/10.1103/PhysRevLett.131.190403} {\bibfield  {journal} {\bibinfo  {journal} {Phys. Rev. Lett.}\ }\textbf {\bibinfo {volume} {131}},\ \bibinfo {pages} {190403} (\bibinfo {year} {2023})}\BibitemShut {NoStop}%
\bibitem [{\citenamefont {Dykman}\ and\ \citenamefont {Smelyanskii}(1988)}]{dykman1988quantum}%
  \BibitemOpen
  \bibfield  {author} {\bibinfo {author} {\bibfnamefont {M.~I.}\ \bibnamefont {Dykman}}\ and\ \bibinfo {author} {\bibfnamefont {V.~N.}\ \bibnamefont {Smelyanskii}},\ }\bibfield  {title} {\bibinfo {title} {Quantum theory of transitions between stable states of a nonlinear oscillator interacting with a medium in a resonant field},\ }\href {https://jetp.ras.ru/cgi-bin/e/index/e/67/9/p1769?a=list} {\bibfield  {journal} {\bibinfo  {journal} {Sov. Phys. JETP}\ }\textbf {\bibinfo {volume} {67}},\ \bibinfo {pages} {1769} (\bibinfo {year} {1988})},\ \bibinfo {note} {originally published in Zh. Eksp. Teor. Fiz. \textbf{94}, 61 (1988)}\BibitemShut {NoStop}%
\bibitem [{\citenamefont {Dykman}(2007)}]{dykman2007critical}%
  \BibitemOpen
  \bibfield  {author} {\bibinfo {author} {\bibfnamefont {M.~I.}\ \bibnamefont {Dykman}},\ }\bibfield  {title} {\bibinfo {title} {Critical exponents in metastable decay via quantum activation},\ }\href {https://doi.org/10.1103/PhysRevE.75.011101} {\bibfield  {journal} {\bibinfo  {journal} {Phys. Rev. E}\ }\textbf {\bibinfo {volume} {75}},\ \bibinfo {pages} {011101} (\bibinfo {year} {2007})}\BibitemShut {NoStop}%
\bibitem [{\citenamefont {Lee}\ \emph {et~al.}(2025)\citenamefont {Lee}, \citenamefont {Brookes}, \citenamefont {Park}, \citenamefont {Szyma\ifmmode~\acute{n}\else \'{n}\fi{}ska},\ and\ \citenamefont {Ginossar}}]{lee2024real}%
  \BibitemOpen
  \bibfield  {author} {\bibinfo {author} {\bibfnamefont {C.-W.}\ \bibnamefont {Lee}}, \bibinfo {author} {\bibfnamefont {P.}~\bibnamefont {Brookes}}, \bibinfo {author} {\bibfnamefont {K.-S.}\ \bibnamefont {Park}}, \bibinfo {author} {\bibfnamefont {M.~H.}\ \bibnamefont {Szyma\ifmmode~\acute{n}\else \'{n}\fi{}ska}},\ and\ \bibinfo {author} {\bibfnamefont {E.}~\bibnamefont {Ginossar}},\ }\bibfield  {title} {\bibinfo {title} {Real-time instanton approach to quantum activation},\ }\href {https://doi.org/10.1103/5jtm-ht4n} {\bibfield  {journal} {\bibinfo  {journal} {Phys. Rev. A}\ }\textbf {\bibinfo {volume} {112}},\ \bibinfo {pages} {012216} (\bibinfo {year} {2025})}\BibitemShut {NoStop}%
\bibitem [{\citenamefont {Xiang}\ \emph {et~al.}(2026)\citenamefont {Xiang}, \citenamefont {Li}, \citenamefont {Bai},\ and\ \citenamefont {Ma}}]{xiang2025switching}%
  \BibitemOpen
  \bibfield  {author} {\bibinfo {author} {\bibfnamefont {Y.-X.}\ \bibnamefont {Xiang}}, \bibinfo {author} {\bibfnamefont {W.}~\bibnamefont {Li}}, \bibinfo {author} {\bibfnamefont {Z.}~\bibnamefont {Bai}},\ and\ \bibinfo {author} {\bibfnamefont {Y.-Q.}\ \bibnamefont {Ma}},\ }\bibfield  {title} {\bibinfo {title} {Switching metastable dynamics in many-body open quantum systems},\ }\href {https://doi.org/10.1093/nsr/nwag146} {\bibfield  {journal} {\bibinfo  {journal} {Natl. Sci. Rev.}\ }\textbf {\bibinfo {volume} {13}},\ \bibinfo {pages} {nwag146} (\bibinfo {year} {2026})}\BibitemShut {NoStop}%
\bibitem [{\citenamefont {S\'epulcre}(2026)}]{sepulcre2026analytical}%
  \BibitemOpen
  \bibfield  {author} {\bibinfo {author} {\bibfnamefont {T.}~\bibnamefont {S\'epulcre}},\ }\bibfield  {title} {\bibinfo {title} {Analytical phase boundary of a quantum driven-dissipative {K}err oscillator from classical stochastic instantons},\ }\href {https://doi.org/10.1103/528d-l76d} {\bibfield  {journal} {\bibinfo  {journal} {Phys. Rev. Res.}\ }\textbf {\bibinfo {volume} {8}},\ \bibinfo {pages} {L012058} (\bibinfo {year} {2026})}\BibitemShut {NoStop}%
\bibitem [{\citenamefont {Mylnikov}\ \emph {et~al.}(2026)\citenamefont {Mylnikov}, \citenamefont {Potashin},\ and\ \citenamefont {Kamenev}}]{mylnikov2026large}%
  \BibitemOpen
  \bibfield  {author} {\bibinfo {author} {\bibfnamefont {V.~Y.}\ \bibnamefont {Mylnikov}}, \bibinfo {author} {\bibfnamefont {S.~O.}\ \bibnamefont {Potashin}},\ and\ \bibinfo {author} {\bibfnamefont {A.}~\bibnamefont {Kamenev}},\ }\href@noop {} {\bibinfo {title} {Large fluctuations in open quantum systems}} (\bibinfo {year} {2026}),\ \Eprint {https://arxiv.org/abs/2606.11822} {arXiv:2606.11822 [quant-ph]} \BibitemShut {NoStop}%
\bibitem [{\citenamefont {Thompson}\ and\ \citenamefont {Kamenev}(2022)}]{thompson2022qubit}%
  \BibitemOpen
  \bibfield  {author} {\bibinfo {author} {\bibfnamefont {F.}~\bibnamefont {Thompson}}\ and\ \bibinfo {author} {\bibfnamefont {A.}~\bibnamefont {Kamenev}},\ }\bibfield  {title} {\bibinfo {title} {Qubit decoherence and symmetry restoration through real-time instantons},\ }\href {https://doi.org/10.1103/PhysRevResearch.4.023020} {\bibfield  {journal} {\bibinfo  {journal} {Phys. Rev. Res.}\ }\textbf {\bibinfo {volume} {4}},\ \bibinfo {pages} {023020} (\bibinfo {year} {2022})}\BibitemShut {NoStop}%
\bibitem [{\citenamefont {Carde}\ \emph {et~al.}(2026)\citenamefont {Carde}, \citenamefont {Gautier}, \citenamefont {Didier}, \citenamefont {Petrescu}, \citenamefont {Cohen},\ and\ \citenamefont {McDonald}}]{carde2026nonperturvative}%
  \BibitemOpen
  \bibfield  {author} {\bibinfo {author} {\bibfnamefont {L.}~\bibnamefont {Carde}}, \bibinfo {author} {\bibfnamefont {R.}~\bibnamefont {Gautier}}, \bibinfo {author} {\bibfnamefont {N.}~\bibnamefont {Didier}}, \bibinfo {author} {\bibfnamefont {A.}~\bibnamefont {Petrescu}}, \bibinfo {author} {\bibfnamefont {J.}~\bibnamefont {Cohen}},\ and\ \bibinfo {author} {\bibfnamefont {A.}~\bibnamefont {McDonald}},\ }\bibfield  {title} {\bibinfo {title} {Nonperturbative switching rates in bistable open quantum systems: From driven {K}err oscillators to dissipative cat qubits},\ }\href {https://doi.org/10.1103/q981-pd5j} {\bibfield  {journal} {\bibinfo  {journal} {Phys. Rev. Lett.}\ }\textbf {\bibinfo {volume} {136}},\ \bibinfo {pages} {100402} (\bibinfo {year} {2026})}\BibitemShut {NoStop}%
\bibitem [{\citenamefont {Mylnikov}\ \emph {et~al.}(2025{\natexlab{a}})\citenamefont {Mylnikov}, \citenamefont {Potashin}, \citenamefont {Ukhtary},\ and\ \citenamefont {Sokolovskii}}]{mylnikov2025switching}%
  \BibitemOpen
  \bibfield  {author} {\bibinfo {author} {\bibfnamefont {V.~Y.}\ \bibnamefont {Mylnikov}}, \bibinfo {author} {\bibfnamefont {S.~O.}\ \bibnamefont {Potashin}}, \bibinfo {author} {\bibfnamefont {M.~S.}\ \bibnamefont {Ukhtary}},\ and\ \bibinfo {author} {\bibfnamefont {G.~S.}\ \bibnamefont {Sokolovskii}},\ }\href@noop {} {\bibinfo {title} {Switching rates in {K}err resonator with two-photon dissipation and driving}} (\bibinfo {year} {2025}{\natexlab{a}}),\ \Eprint {https://arxiv.org/abs/2511.13308} {arXiv:2511.13308 [quant-ph]} \BibitemShut {NoStop}%
\bibitem [{\citenamefont {Mylnikov}\ \emph {et~al.}(2025{\natexlab{b}})\citenamefont {Mylnikov}, \citenamefont {Potashin},\ and\ \citenamefont {Kamenev}}]{mylnikov2025qubit}%
  \BibitemOpen
  \bibfield  {author} {\bibinfo {author} {\bibfnamefont {V.~Y.}\ \bibnamefont {Mylnikov}}, \bibinfo {author} {\bibfnamefont {S.~O.}\ \bibnamefont {Potashin}},\ and\ \bibinfo {author} {\bibfnamefont {A.}~\bibnamefont {Kamenev}},\ }\href@noop {} {\bibinfo {title} {Qubit decoherence in dissipative two-photon resonator: real-time instantons and {W}igner function}} (\bibinfo {year} {2025}{\natexlab{b}}),\ \Eprint {https://arxiv.org/abs/2512.10921} {arXiv:2512.10921 [quant-ph]} \BibitemShut {NoStop}%
\bibitem [{\citenamefont {Thompson}\ \emph {et~al.}(2026)\citenamefont {Thompson}, \citenamefont {Bone\ss{}}, \citenamefont {Dykman},\ and\ \citenamefont {Kamenev}}]{thompson2026spectroscopy}%
  \BibitemOpen
  \bibfield  {author} {\bibinfo {author} {\bibfnamefont {F.}~\bibnamefont {Thompson}}, \bibinfo {author} {\bibfnamefont {D.~K.~J.}\ \bibnamefont {Bone\ss{}}}, \bibinfo {author} {\bibfnamefont {M.}~\bibnamefont {Dykman}},\ and\ \bibinfo {author} {\bibfnamefont {A.}~\bibnamefont {Kamenev}},\ }\bibfield  {title} {\bibinfo {title} {Spectroscopy of quantum phase slips: Visualizing complex real-time instantons},\ }\href {https://doi.org/10.1103/bkyx-wwv1} {\bibfield  {journal} {\bibinfo  {journal} {Phys. Rev. A}\ }\textbf {\bibinfo {volume} {113}},\ \bibinfo {pages} {043712} (\bibinfo {year} {2026})}\BibitemShut {NoStop}%
\bibitem [{\citenamefont {Mivehvar}\ \emph {et~al.}(2021)\citenamefont {Mivehvar}, \citenamefont {Piazza}, \citenamefont {Donner},\ and\ \citenamefont {Ritsch}}]{mivehvar2021cavity}%
  \BibitemOpen
  \bibfield  {author} {\bibinfo {author} {\bibfnamefont {F.}~\bibnamefont {Mivehvar}}, \bibinfo {author} {\bibfnamefont {F.}~\bibnamefont {Piazza}}, \bibinfo {author} {\bibfnamefont {T.}~\bibnamefont {Donner}},\ and\ \bibinfo {author} {\bibfnamefont {H.}~\bibnamefont {Ritsch}},\ }\bibfield  {title} {\bibinfo {title} {Cavity {QED} with quantum gases: New paradigms in many-body physics},\ }\href {https://doi.org/10.1080/00018732.2021.1969727} {\bibfield  {journal} {\bibinfo  {journal} {Advances in Physics}\ }\textbf {\bibinfo {volume} {70}},\ \bibinfo {pages} {1} (\bibinfo {year} {2021})}\BibitemShut {NoStop}%
\bibitem [{\citenamefont {Morrison}\ and\ \citenamefont {Parkins}(2008{\natexlab{a}})}]{morrison2008dynamicalprl}%
  \BibitemOpen
  \bibfield  {author} {\bibinfo {author} {\bibfnamefont {S.}~\bibnamefont {Morrison}}\ and\ \bibinfo {author} {\bibfnamefont {A.~S.}\ \bibnamefont {Parkins}},\ }\bibfield  {title} {\bibinfo {title} {Dynamical quantum phase transitions in the dissipative {Lipkin-Meshkov-Glick} model with proposed realization in optical cavity {QED}},\ }\href {https://doi.org/10.1103/PhysRevLett.100.040403} {\bibfield  {journal} {\bibinfo  {journal} {Phys. Rev. Lett.}\ }\textbf {\bibinfo {volume} {100}},\ \bibinfo {pages} {040403} (\bibinfo {year} {2008}{\natexlab{a}})}\BibitemShut {NoStop}%
\bibitem [{\citenamefont {Morrison}\ and\ \citenamefont {Parkins}(2008{\natexlab{b}})}]{morrison2008collective}%
  \BibitemOpen
  \bibfield  {author} {\bibinfo {author} {\bibfnamefont {S.}~\bibnamefont {Morrison}}\ and\ \bibinfo {author} {\bibfnamefont {A.~S.}\ \bibnamefont {Parkins}},\ }\bibfield  {title} {\bibinfo {title} {Collective spin systems in dispersive optical cavity {QED}: Quantum phase transitions and entanglement},\ }\href {https://doi.org/10.1103/PhysRevA.77.043810} {\bibfield  {journal} {\bibinfo  {journal} {Phys. Rev. A}\ }\textbf {\bibinfo {volume} {77}},\ \bibinfo {pages} {043810} (\bibinfo {year} {2008}{\natexlab{b}})}\BibitemShut {NoStop}%
\bibitem [{\citenamefont {Norcia}\ \emph {et~al.}(2018)\citenamefont {Norcia}, \citenamefont {Lewis-Swan}, \citenamefont {Cline}, \citenamefont {Zhu}, \citenamefont {Rey},\ and\ \citenamefont {Thompson}}]{norcia2018cavity}%
  \BibitemOpen
  \bibfield  {author} {\bibinfo {author} {\bibfnamefont {M.~A.}\ \bibnamefont {Norcia}}, \bibinfo {author} {\bibfnamefont {R.~J.}\ \bibnamefont {Lewis-Swan}}, \bibinfo {author} {\bibfnamefont {J.~R.}\ \bibnamefont {Cline}}, \bibinfo {author} {\bibfnamefont {B.}~\bibnamefont {Zhu}}, \bibinfo {author} {\bibfnamefont {A.~M.}\ \bibnamefont {Rey}},\ and\ \bibinfo {author} {\bibfnamefont {J.~K.}\ \bibnamefont {Thompson}},\ }\bibfield  {title} {\bibinfo {title} {Cavity-mediated collective spin-exchange interactions in a strontium superradiant laser},\ }\href {https://doi.org/10.1126/science.aar3102} {\bibfield  {journal} {\bibinfo  {journal} {Science}\ }\textbf {\bibinfo {volume} {361}},\ \bibinfo {pages} {259} (\bibinfo {year} {2018})}\BibitemShut {NoStop}%
\bibitem [{\citenamefont {Muniz}\ \emph {et~al.}(2020)\citenamefont {Muniz}, \citenamefont {Barberena}, \citenamefont {Lewis-Swan}, \citenamefont {Young}, \citenamefont {Cline}, \citenamefont {Rey},\ and\ \citenamefont {Thompson}}]{muniz2020exploring}%
  \BibitemOpen
  \bibfield  {author} {\bibinfo {author} {\bibfnamefont {J.~A.}\ \bibnamefont {Muniz}}, \bibinfo {author} {\bibfnamefont {D.}~\bibnamefont {Barberena}}, \bibinfo {author} {\bibfnamefont {R.~J.}\ \bibnamefont {Lewis-Swan}}, \bibinfo {author} {\bibfnamefont {D.~J.}\ \bibnamefont {Young}}, \bibinfo {author} {\bibfnamefont {J.~R.}\ \bibnamefont {Cline}}, \bibinfo {author} {\bibfnamefont {A.~M.}\ \bibnamefont {Rey}},\ and\ \bibinfo {author} {\bibfnamefont {J.~K.}\ \bibnamefont {Thompson}},\ }\bibfield  {title} {\bibinfo {title} {Exploring dynamical phase transitions with cold atoms in an optical cavity},\ }\href {https://doi.org/10.1038/s41586-020-2224-x} {\bibfield  {journal} {\bibinfo  {journal} {Nature}\ }\textbf {\bibinfo {volume} {580}},\ \bibinfo {pages} {602} (\bibinfo {year} {2020})}\BibitemShut {NoStop}%
\bibitem [{\citenamefont {Song}\ \emph {et~al.}(2025)\citenamefont {Song}, \citenamefont {Barberena}, \citenamefont {Young}, \citenamefont {Chaparro}, \citenamefont {Chu}, \citenamefont {Agarwal}, \citenamefont {Niu}, \citenamefont {Young}, \citenamefont {Rey},\ and\ \citenamefont {Thompson}}]{song2025dissipation}%
  \BibitemOpen
  \bibfield  {author} {\bibinfo {author} {\bibfnamefont {E.~Y.}\ \bibnamefont {Song}}, \bibinfo {author} {\bibfnamefont {D.}~\bibnamefont {Barberena}}, \bibinfo {author} {\bibfnamefont {D.~J.}\ \bibnamefont {Young}}, \bibinfo {author} {\bibfnamefont {E.}~\bibnamefont {Chaparro}}, \bibinfo {author} {\bibfnamefont {A.}~\bibnamefont {Chu}}, \bibinfo {author} {\bibfnamefont {S.}~\bibnamefont {Agarwal}}, \bibinfo {author} {\bibfnamefont {Z.}~\bibnamefont {Niu}}, \bibinfo {author} {\bibfnamefont {J.~T.}\ \bibnamefont {Young}}, \bibinfo {author} {\bibfnamefont {A.~M.}\ \bibnamefont {Rey}},\ and\ \bibinfo {author} {\bibfnamefont {J.~K.}\ \bibnamefont {Thompson}},\ }\bibfield  {title} {\bibinfo {title} {{A dissipation-induced superradiant transition in a strontium cavity-QED system}},\ }\href {https://doi.org/10.1126/sciadv.adu5799} {\bibfield  {journal} {\bibinfo  {journal} {Sci. Adv.}\ }\textbf {\bibinfo {volume} {11}},\ \bibinfo {pages} {eadu5799} (\bibinfo {year} {2025})}\BibitemShut {NoStop}%
\bibitem [{\citenamefont {Ferri}\ \emph {et~al.}(2021)\citenamefont {Ferri}, \citenamefont {Rosa-Medina}, \citenamefont {Finger}, \citenamefont {Dogra}, \citenamefont {Soriente}, \citenamefont {Zilberberg}, \citenamefont {Donner},\ and\ \citenamefont {Esslinger}}]{ferri2021emerging}%
  \BibitemOpen
  \bibfield  {author} {\bibinfo {author} {\bibfnamefont {F.}~\bibnamefont {Ferri}}, \bibinfo {author} {\bibfnamefont {R.}~\bibnamefont {Rosa-Medina}}, \bibinfo {author} {\bibfnamefont {F.}~\bibnamefont {Finger}}, \bibinfo {author} {\bibfnamefont {N.}~\bibnamefont {Dogra}}, \bibinfo {author} {\bibfnamefont {M.}~\bibnamefont {Soriente}}, \bibinfo {author} {\bibfnamefont {O.}~\bibnamefont {Zilberberg}}, \bibinfo {author} {\bibfnamefont {T.}~\bibnamefont {Donner}},\ and\ \bibinfo {author} {\bibfnamefont {T.}~\bibnamefont {Esslinger}},\ }\bibfield  {title} {\bibinfo {title} {Emerging dissipative phases in a superradiant quantum gas with tunable decay},\ }\href {https://doi.org/10.1103/PhysRevX.11.041046} {\bibfield  {journal} {\bibinfo  {journal} {Phys. Rev. X}\ }\textbf {\bibinfo {volume} {11}},\ \bibinfo {pages} {041046} (\bibinfo {year} {2021})}\BibitemShut {NoStop}%
\bibitem [{\citenamefont {Mlynek}\ \emph {et~al.}(2012)\citenamefont {Mlynek}, \citenamefont {Abdumalikov}, \citenamefont {Fink}, \citenamefont {Steffen}, \citenamefont {Baur}, \citenamefont {Lang}, \citenamefont {van Loo},\ and\ \citenamefont {Wallraff}}]{mlynek2012demonstrating}%
  \BibitemOpen
  \bibfield  {author} {\bibinfo {author} {\bibfnamefont {J.~A.}\ \bibnamefont {Mlynek}}, \bibinfo {author} {\bibfnamefont {A.~A.}\ \bibnamefont {Abdumalikov}}, \bibinfo {author} {\bibfnamefont {J.~M.}\ \bibnamefont {Fink}}, \bibinfo {author} {\bibfnamefont {L.}~\bibnamefont {Steffen}}, \bibinfo {author} {\bibfnamefont {M.}~\bibnamefont {Baur}}, \bibinfo {author} {\bibfnamefont {C.}~\bibnamefont {Lang}}, \bibinfo {author} {\bibfnamefont {A.~F.}\ \bibnamefont {van Loo}},\ and\ \bibinfo {author} {\bibfnamefont {A.}~\bibnamefont {Wallraff}},\ }\bibfield  {title} {\bibinfo {title} {Demonstrating {$W$}-type entanglement of {D}icke states in resonant cavity quantum electrodynamics},\ }\href {https://doi.org/10.1103/PhysRevA.86.053838} {\bibfield  {journal} {\bibinfo  {journal} {Phys. Rev. A}\ }\textbf {\bibinfo {volume} {86}},\ \bibinfo {pages} {053838} (\bibinfo {year} {2012})}\BibitemShut {NoStop}%
\bibitem [{\citenamefont {Nissen}\ \emph {et~al.}(2013)\citenamefont {Nissen}, \citenamefont {Fink}, \citenamefont {Mlynek}, \citenamefont {Wallraff},\ and\ \citenamefont {Keeling}}]{nissen2013collective}%
  \BibitemOpen
  \bibfield  {author} {\bibinfo {author} {\bibfnamefont {F.}~\bibnamefont {Nissen}}, \bibinfo {author} {\bibfnamefont {J.~M.}\ \bibnamefont {Fink}}, \bibinfo {author} {\bibfnamefont {J.~A.}\ \bibnamefont {Mlynek}}, \bibinfo {author} {\bibfnamefont {A.}~\bibnamefont {Wallraff}},\ and\ \bibinfo {author} {\bibfnamefont {J.}~\bibnamefont {Keeling}},\ }\bibfield  {title} {\bibinfo {title} {Collective suppression of linewidths in circuit {QED}},\ }\href {https://doi.org/10.1103/PhysRevLett.110.203602} {\bibfield  {journal} {\bibinfo  {journal} {Phys. Rev. Lett.}\ }\textbf {\bibinfo {volume} {110}},\ \bibinfo {pages} {203602} (\bibinfo {year} {2013})}\BibitemShut {NoStop}%
\bibitem [{\citenamefont {Kessler}\ \emph {et~al.}(2012)\citenamefont {Kessler}, \citenamefont {Giedke}, \citenamefont {Imamoglu}, \citenamefont {Yelin}, \citenamefont {Lukin},\ and\ \citenamefont {Cirac}}]{kessler2012dissipative}%
  \BibitemOpen
  \bibfield  {author} {\bibinfo {author} {\bibfnamefont {E.~M.}\ \bibnamefont {Kessler}}, \bibinfo {author} {\bibfnamefont {G.}~\bibnamefont {Giedke}}, \bibinfo {author} {\bibfnamefont {A.}~\bibnamefont {Imamoglu}}, \bibinfo {author} {\bibfnamefont {S.~F.}\ \bibnamefont {Yelin}}, \bibinfo {author} {\bibfnamefont {M.~D.}\ \bibnamefont {Lukin}},\ and\ \bibinfo {author} {\bibfnamefont {J.~I.}\ \bibnamefont {Cirac}},\ }\bibfield  {title} {\bibinfo {title} {Dissipative phase transition in a central spin system},\ }\href {https://doi.org/10.1103/PhysRevA.86.012116} {\bibfield  {journal} {\bibinfo  {journal} {Phys. Rev. A}\ }\textbf {\bibinfo {volume} {86}},\ \bibinfo {pages} {012116} (\bibinfo {year} {2012})}\BibitemShut {NoStop}%
\bibitem [{\citenamefont {Shakirov}\ \emph {et~al.}(2016)\citenamefont {Shakirov}, \citenamefont {Shchadilova}, \citenamefont {Rubtsov},\ and\ \citenamefont {Ribeiro}}]{shakirov2016role}%
  \BibitemOpen
  \bibfield  {author} {\bibinfo {author} {\bibfnamefont {A.~M.}\ \bibnamefont {Shakirov}}, \bibinfo {author} {\bibfnamefont {Y.~E.}\ \bibnamefont {Shchadilova}}, \bibinfo {author} {\bibfnamefont {A.~N.}\ \bibnamefont {Rubtsov}},\ and\ \bibinfo {author} {\bibfnamefont {P.}~\bibnamefont {Ribeiro}},\ }\bibfield  {title} {\bibinfo {title} {Role of coherence in transport through engineered atomic spin devices},\ }\href {https://doi.org/10.1103/PhysRevB.94.224425} {\bibfield  {journal} {\bibinfo  {journal} {Phys. Rev. B}\ }\textbf {\bibinfo {volume} {94}},\ \bibinfo {pages} {224425} (\bibinfo {year} {2016})}\BibitemShut {NoStop}%
\bibitem [{\citenamefont {Ferreira}\ and\ \citenamefont {Ribeiro}(2019)}]{ferreira2019lipkin}%
  \BibitemOpen
  \bibfield  {author} {\bibinfo {author} {\bibfnamefont {J.~S.}\ \bibnamefont {Ferreira}}\ and\ \bibinfo {author} {\bibfnamefont {P.}~\bibnamefont {Ribeiro}},\ }\bibfield  {title} {\bibinfo {title} {Lipkin-{M}eshkov-{G}lick model with {M}arkovian dissipation: A description of a collective spin on a metallic surface},\ }\href {https://doi.org/10.1103/PhysRevB.100.184422} {\bibfield  {journal} {\bibinfo  {journal} {Phys. Rev. B}\ }\textbf {\bibinfo {volume} {100}},\ \bibinfo {pages} {184422} (\bibinfo {year} {2019})}\BibitemShut {NoStop}%
\bibitem [{\citenamefont {Kiselev}\ and\ \citenamefont {Oppermann}(2000)}]{kiselev2000schwinger}%
  \BibitemOpen
  \bibfield  {author} {\bibinfo {author} {\bibfnamefont {M.~N.}\ \bibnamefont {Kiselev}}\ and\ \bibinfo {author} {\bibfnamefont {R.}~\bibnamefont {Oppermann}},\ }\bibfield  {title} {\bibinfo {title} {{Schwinger-Keldysh Semionic Approach for Quantum Spin Systems}},\ }\href {https://doi.org/10.1103/PhysRevLett.85.5631} {\bibfield  {journal} {\bibinfo  {journal} {Phys. Rev. Lett.}\ }\textbf {\bibinfo {volume} {85}},\ \bibinfo {pages} {5631} (\bibinfo {year} {2000})}\BibitemShut {NoStop}%
\bibitem [{\citenamefont {Dutta}\ \emph {et~al.}(2025)\citenamefont {Dutta}, \citenamefont {Zhang},\ and\ \citenamefont {Haque}}]{dutta2025quantum}%
  \BibitemOpen
  \bibfield  {author} {\bibinfo {author} {\bibfnamefont {S.}~\bibnamefont {Dutta}}, \bibinfo {author} {\bibfnamefont {S.}~\bibnamefont {Zhang}},\ and\ \bibinfo {author} {\bibfnamefont {M.}~\bibnamefont {Haque}},\ }\bibfield  {title} {\bibinfo {title} {{Quantum Origin of Limit Cycles, Fixed Points, and Critical Slowing Down}},\ }\href {https://doi.org/10.1103/PhysRevLett.134.050407} {\bibfield  {journal} {\bibinfo  {journal} {Phys. Rev. Lett.}\ }\textbf {\bibinfo {volume} {134}},\ \bibinfo {pages} {050407} (\bibinfo {year} {2025})}\BibitemShut {NoStop}%
\bibitem [{\citenamefont {Maier}\ and\ \citenamefont {Stein}(1993)}]{meier1993escape}%
  \BibitemOpen
  \bibfield  {author} {\bibinfo {author} {\bibfnamefont {R.~S.}\ \bibnamefont {Maier}}\ and\ \bibinfo {author} {\bibfnamefont {D.~L.}\ \bibnamefont {Stein}},\ }\bibfield  {title} {\bibinfo {title} {Escape problem for irreversible systems},\ }\href {https://doi.org/10.1103/PhysRevE.48.931} {\bibfield  {journal} {\bibinfo  {journal} {Phys. Rev. E}\ }\textbf {\bibinfo {volume} {48}},\ \bibinfo {pages} {931} (\bibinfo {year} {1993})}\BibitemShut {NoStop}%
\bibitem [{\citenamefont {Dykman}\ \emph {et~al.}(1994)\citenamefont {Dykman}, \citenamefont {Mori}, \citenamefont {Ross},\ and\ \citenamefont {Hunt}}]{dykman1994large}%
  \BibitemOpen
  \bibfield  {author} {\bibinfo {author} {\bibfnamefont {M.~I.}\ \bibnamefont {Dykman}}, \bibinfo {author} {\bibfnamefont {E.}~\bibnamefont {Mori}}, \bibinfo {author} {\bibfnamefont {J.}~\bibnamefont {Ross}},\ and\ \bibinfo {author} {\bibfnamefont {P.}~\bibnamefont {Hunt}},\ }\bibfield  {title} {\bibinfo {title} {Large fluctuations and optimal paths in chemical kinetics},\ }\href {https://doi.org/10.1063/1.467139} {\bibfield  {journal} {\bibinfo  {journal} {J. Chem. Phys.}\ }\textbf {\bibinfo {volume} {100}},\ \bibinfo {pages} {5735} (\bibinfo {year} {1994})}\BibitemShut {NoStop}%
\bibitem [{\citenamefont {Grafke}\ and\ \citenamefont {Vanden-Eijnden}(2019)}]{grafke2019numerical}%
  \BibitemOpen
  \bibfield  {author} {\bibinfo {author} {\bibfnamefont {T.}~\bibnamefont {Grafke}}\ and\ \bibinfo {author} {\bibfnamefont {E.}~\bibnamefont {Vanden-Eijnden}},\ }\bibfield  {title} {\bibinfo {title} {Numerical computation of rare events via large deviation theory},\ }\href {https://doi.org/10.1063/1.5084025} {\bibfield  {journal} {\bibinfo  {journal} {Chaos}\ }\textbf {\bibinfo {volume} {29}},\ \bibinfo {pages} {063118} (\bibinfo {year} {2019})}\BibitemShut {NoStop}%
\bibitem [{\citenamefont {Gagrani}\ and\ \citenamefont {Smith}(2023)}]{gagrani2023action}%
  \BibitemOpen
  \bibfield  {author} {\bibinfo {author} {\bibfnamefont {P.}~\bibnamefont {Gagrani}}\ and\ \bibinfo {author} {\bibfnamefont {E.}~\bibnamefont {Smith}},\ }\bibfield  {title} {\bibinfo {title} {Action functional gradient descent algorithm for estimating escape paths in stochastic chemical reaction networks},\ }\href {https://doi.org/10.1103/PhysRevE.107.034305} {\bibfield  {journal} {\bibinfo  {journal} {Phys. Rev. E}\ }\textbf {\bibinfo {volume} {107}},\ \bibinfo {pages} {034305} (\bibinfo {year} {2023})}\BibitemShut {NoStop}%
\bibitem [{\citenamefont {Zakine}\ and\ \citenamefont {Vanden-Eijnden}(2023)}]{zakine2023minimum}%
  \BibitemOpen
  \bibfield  {author} {\bibinfo {author} {\bibfnamefont {R.}~\bibnamefont {Zakine}}\ and\ \bibinfo {author} {\bibfnamefont {E.}~\bibnamefont {Vanden-Eijnden}},\ }\bibfield  {title} {\bibinfo {title} {{Minimum-Action Method for Nonequilibrium Phase Transitions}},\ }\href {https://doi.org/10.1103/PhysRevX.13.041044} {\bibfield  {journal} {\bibinfo  {journal} {Phys. Rev. X}\ }\textbf {\bibinfo {volume} {13}},\ \bibinfo {pages} {041044} (\bibinfo {year} {2023})}\BibitemShut {NoStop}%
\bibitem [{\citenamefont {Falasco}\ and\ \citenamefont {Esposito}(2025)}]{FalascoReview}%
  \BibitemOpen
  \bibfield  {author} {\bibinfo {author} {\bibfnamefont {G.}~\bibnamefont {Falasco}}\ and\ \bibinfo {author} {\bibfnamefont {M.}~\bibnamefont {Esposito}},\ }\bibfield  {title} {\bibinfo {title} {Macroscopic stochastic thermodynamics},\ }\href {https://doi.org/10.1103/RevModPhys.97.015002} {\bibfield  {journal} {\bibinfo  {journal} {Rev. Mod. Phys.}\ }\textbf {\bibinfo {volume} {97}},\ \bibinfo {pages} {015002} (\bibinfo {year} {2025})}\BibitemShut {NoStop}%
\bibitem [{\citenamefont {Chudak}\ \emph {et~al.}(2026)\citenamefont {Chudak}, \citenamefont {Esposito},\ and\ \citenamefont {Ptaszynski}}]{chudak2026quantum}%
  \BibitemOpen
  \bibfield  {author} {\bibinfo {author} {\bibfnamefont {M.}~\bibnamefont {Chudak}}, \bibinfo {author} {\bibfnamefont {M.}~\bibnamefont {Esposito}},\ and\ \bibinfo {author} {\bibfnamefont {K.}~\bibnamefont {Ptaszynski}},\ }\href@noop {} {\bibinfo {title} {Quantum instanton approach to metastable collective spin systems: Numerical implementation and applications}} (\bibinfo {year} {2026}),\ \Eprint {https://arxiv.org/abs/2610.21684} {arXiv:2610.21684 [quant-ph]} \BibitemShut {NoStop}%
\bibitem [{\citenamefont {Shammah}\ \emph {et~al.}(2018)\citenamefont {Shammah}, \citenamefont {Ahmed}, \citenamefont {Lambert}, \citenamefont {De~Liberato},\ and\ \citenamefont {Nori}}]{shammah2018open}%
  \BibitemOpen
  \bibfield  {author} {\bibinfo {author} {\bibfnamefont {N.}~\bibnamefont {Shammah}}, \bibinfo {author} {\bibfnamefont {S.}~\bibnamefont {Ahmed}}, \bibinfo {author} {\bibfnamefont {N.}~\bibnamefont {Lambert}}, \bibinfo {author} {\bibfnamefont {S.}~\bibnamefont {De~Liberato}},\ and\ \bibinfo {author} {\bibfnamefont {F.}~\bibnamefont {Nori}},\ }\bibfield  {title} {\bibinfo {title} {Open quantum systems with local and collective incoherent processes: Efficient numerical simulations using permutational invariance},\ }\href {https://doi.org/10.1103/PhysRevA.98.063815} {\bibfield  {journal} {\bibinfo  {journal} {Phys. Rev. A}\ }\textbf {\bibinfo {volume} {98}},\ \bibinfo {pages} {063815} (\bibinfo {year} {2018})}\BibitemShut {NoStop}%
\bibitem [{sup()}]{supp}%
  \BibitemOpen
  \href@noop {} {}\bibinfo {note} {{See Supplemental Material at ($\ldots$), which includes Refs.~\cite{nation2015iterative,dubois2021semi}, for calculation details, description of the SW approach, and additional results for $\Omega=0.5\gamma$.}}\BibitemShut {Stop}%
\bibitem [{\citenamefont {Keizer}(1978)}]{keizer1978thermodynamics}%
  \BibitemOpen
  \bibfield  {author} {\bibinfo {author} {\bibfnamefont {J.}~\bibnamefont {Keizer}},\ }\bibfield  {title} {\bibinfo {title} {Thermodynamics at nonequilibrium steady states},\ }\href {https://doi.org/10.1063/1.436908} {\bibfield  {journal} {\bibinfo  {journal} {J. Chem. Phys.}\ }\textbf {\bibinfo {volume} {69}},\ \bibinfo {pages} {2609} (\bibinfo {year} {1978})}\BibitemShut {NoStop}%
\bibitem [{\citenamefont {Narducci}\ \emph {et~al.}(1975)\citenamefont {Narducci}, \citenamefont {Bowden}, \citenamefont {Bluemel}, \citenamefont {Garrazana},\ and\ \citenamefont {Tuft}}]{narducci1975multitime}%
  \BibitemOpen
  \bibfield  {author} {\bibinfo {author} {\bibfnamefont {L.~M.}\ \bibnamefont {Narducci}}, \bibinfo {author} {\bibfnamefont {C.~M.}\ \bibnamefont {Bowden}}, \bibinfo {author} {\bibfnamefont {V.}~\bibnamefont {Bluemel}}, \bibinfo {author} {\bibfnamefont {G.~P.}\ \bibnamefont {Garrazana}},\ and\ \bibinfo {author} {\bibfnamefont {R.~A.}\ \bibnamefont {Tuft}},\ }\bibfield  {title} {\bibinfo {title} {Multitime-correlation functions and the atomic coherent-state representation},\ }\href {https://doi.org/10.1103/PhysRevA.11.973} {\bibfield  {journal} {\bibinfo  {journal} {Phys. Rev. A}\ }\textbf {\bibinfo {volume} {11}},\ \bibinfo {pages} {973} (\bibinfo {year} {1975})}\BibitemShut {NoStop}%
\bibitem [{\citenamefont {Altland}\ and\ \citenamefont {Haake}(2012)}]{altland2012quantum}%
  \BibitemOpen
  \bibfield  {author} {\bibinfo {author} {\bibfnamefont {A.}~\bibnamefont {Altland}}\ and\ \bibinfo {author} {\bibfnamefont {F.}~\bibnamefont {Haake}},\ }\bibfield  {title} {\bibinfo {title} {Quantum chaos and effective thermalization},\ }\href {https://doi.org/10.1103/PhysRevLett.108.073601} {\bibfield  {journal} {\bibinfo  {journal} {Phys. Rev. Lett.}\ }\textbf {\bibinfo {volume} {108}},\ \bibinfo {pages} {073601} (\bibinfo {year} {2012})}\BibitemShut {NoStop}%
\bibitem [{\citenamefont {Mandt}\ \emph {et~al.}(2015)\citenamefont {Mandt}, \citenamefont {Sadri}, \citenamefont {Houck},\ and\ \citenamefont {T{\"u}reci}}]{mandt2015stochastic}%
  \BibitemOpen
  \bibfield  {author} {\bibinfo {author} {\bibfnamefont {S.}~\bibnamefont {Mandt}}, \bibinfo {author} {\bibfnamefont {D.}~\bibnamefont {Sadri}}, \bibinfo {author} {\bibfnamefont {A.~A.}\ \bibnamefont {Houck}},\ and\ \bibinfo {author} {\bibfnamefont {H.~E.}\ \bibnamefont {T{\"u}reci}},\ }\bibfield  {title} {\bibinfo {title} {Stochastic differential equations for quantum dynamics of spin-boson networks},\ }\href {https://doi.org/10.1088/1367-2630/17/5/053018} {\bibfield  {journal} {\bibinfo  {journal} {New J. Phys.}\ }\textbf {\bibinfo {volume} {17}},\ \bibinfo {pages} {053018} (\bibinfo {year} {2015})}\BibitemShut {NoStop}%
\bibitem [{\citenamefont {B\"urkle}\ and\ \citenamefont {Anglin}(2020)}]{burkle2020probabilistic}%
  \BibitemOpen
  \bibfield  {author} {\bibinfo {author} {\bibfnamefont {R.}~\bibnamefont {B\"urkle}}\ and\ \bibinfo {author} {\bibfnamefont {J.~R.}\ \bibnamefont {Anglin}},\ }\bibfield  {title} {\bibinfo {title} {Probabilistic hysteresis from a quantum-phase-space perspective},\ }\href {https://doi.org/10.1103/PhysRevA.102.052212} {\bibfield  {journal} {\bibinfo  {journal} {Phys. Rev. A}\ }\textbf {\bibinfo {volume} {102}},\ \bibinfo {pages} {052212} (\bibinfo {year} {2020})}\BibitemShut {NoStop}%
\bibitem [{\citenamefont {Radcliffe}(1971)}]{radcliffe1971some}%
  \BibitemOpen
  \bibfield  {author} {\bibinfo {author} {\bibfnamefont {J.~M.}\ \bibnamefont {Radcliffe}},\ }\bibfield  {title} {\bibinfo {title} {Some properties of coherent spin states},\ }\href {https://doi.org/10.1088/0305-4470/4/3/009} {\bibfield  {journal} {\bibinfo  {journal} {Journal of Physics A: General Physics}\ }\textbf {\bibinfo {volume} {4}},\ \bibinfo {pages} {313} (\bibinfo {year} {1971})}\BibitemShut {NoStop}%
\bibitem [{\citenamefont {Bertini}\ \emph {et~al.}(2010)\citenamefont {Bertini}, \citenamefont {Sole}, \citenamefont {Gabrielli}, \citenamefont {Jona-Lasinio},\ and\ \citenamefont {Landim}}]{bertini2010lagrangian}%
  \BibitemOpen
  \bibfield  {author} {\bibinfo {author} {\bibfnamefont {L.}~\bibnamefont {Bertini}}, \bibinfo {author} {\bibfnamefont {A.~D.}\ \bibnamefont {Sole}}, \bibinfo {author} {\bibfnamefont {D.}~\bibnamefont {Gabrielli}}, \bibinfo {author} {\bibfnamefont {G.}~\bibnamefont {Jona-Lasinio}},\ and\ \bibinfo {author} {\bibfnamefont {C.}~\bibnamefont {Landim}},\ }\bibfield  {title} {\bibinfo {title} {Lagrangian phase transitions in nonequilibrium thermodynamic systems},\ }\href {https://doi.org/10.1088/1742-5468/2010/11/L11001} {\bibfield  {journal} {\bibinfo  {journal} {Journal of Statistical Mechanics: Theory and Experiment}\ }\textbf {\bibinfo {volume} {2010}},\ \bibinfo {pages} {L11001} (\bibinfo {year} {2010})}\BibitemShut {NoStop}%
\bibitem [{\citenamefont {Carmichael}(1999)}]{carmichael1999statistical1}%
  \BibitemOpen
  \bibfield  {author} {\bibinfo {author} {\bibfnamefont {H.~J.}\ \bibnamefont {Carmichael}},\ }\href {https://doi.org/10.1007/978-3-662-03875-8} {\emph {\bibinfo {title} {{Statistical Methods in Quantum Optics 1}}}}\ (\bibinfo  {publisher} {Springer},\ \bibinfo {address} {Berlin},\ \bibinfo {year} {1999})\BibitemShut {NoStop}%
\bibitem [{\citenamefont {H\"{a}nggi}\ and\ \citenamefont {Jung}(1988)}]{hanggi1988bistability}%
  \BibitemOpen
  \bibfield  {author} {\bibinfo {author} {\bibfnamefont {P.}~\bibnamefont {H\"{a}nggi}}\ and\ \bibinfo {author} {\bibfnamefont {P.}~\bibnamefont {Jung}},\ }\bibfield  {title} {\bibinfo {title} {Bistability in active circuits: Application of a novel {F}okker-{P}lanck approach},\ }\href {https://doi.org/10.1147/rd.321.0119} {\bibfield  {journal} {\bibinfo  {journal} {IBM J. Res. Dev.}\ }\textbf {\bibinfo {volume} {32}},\ \bibinfo {pages} {119} (\bibinfo {year} {1988})}\BibitemShut {NoStop}%
\bibitem [{\citenamefont {Gaveau}\ \emph {et~al.}(1997)\citenamefont {Gaveau}, \citenamefont {Moreau},\ and\ \citenamefont {Toth}}]{gaveau1997master}%
  \BibitemOpen
  \bibfield  {author} {\bibinfo {author} {\bibfnamefont {B.}~\bibnamefont {Gaveau}}, \bibinfo {author} {\bibfnamefont {M.}~\bibnamefont {Moreau}},\ and\ \bibinfo {author} {\bibfnamefont {J.}~\bibnamefont {Toth}},\ }\bibfield  {title} {\bibinfo {title} {Master equation and {F}okker--{P}lanck equation: comparison of entropy and of rate constants},\ }\href {https://doi.org/10.1023/A:1007362811930} {\bibfield  {journal} {\bibinfo  {journal} {Lett. Math. Phys.}\ }\textbf {\bibinfo {volume} {40}},\ \bibinfo {pages} {101} (\bibinfo {year} {1997})}\BibitemShut {NoStop}%
\bibitem [{\citenamefont {Kessler}\ and\ \citenamefont {Shnerb}(2007)}]{kessler2007extinction}%
  \BibitemOpen
  \bibfield  {author} {\bibinfo {author} {\bibfnamefont {D.~A.}\ \bibnamefont {Kessler}}\ and\ \bibinfo {author} {\bibfnamefont {N.~M.}\ \bibnamefont {Shnerb}},\ }\bibfield  {title} {\bibinfo {title} {Extinction rates for fluctuation-induced metastabilities: A real-space {WKB} approach},\ }\href {https://doi.org/10.1007/s10955-007-9312-2} {\bibfield  {journal} {\bibinfo  {journal} {J. Stat. Phys.}\ }\textbf {\bibinfo {volume} {127}},\ \bibinfo {pages} {861} (\bibinfo {year} {2007})}\BibitemShut {NoStop}%
\bibitem [{\citenamefont {Debecker}\ \emph {et~al.}(2024)\citenamefont {Debecker}, \citenamefont {Martin},\ and\ \citenamefont {Damanet}}]{debecker2024controlling}%
  \BibitemOpen
  \bibfield  {author} {\bibinfo {author} {\bibfnamefont {B.}~\bibnamefont {Debecker}}, \bibinfo {author} {\bibfnamefont {J.}~\bibnamefont {Martin}},\ and\ \bibinfo {author} {\bibfnamefont {F.}~\bibnamefont {Damanet}},\ }\bibfield  {title} {\bibinfo {title} {Controlling matter phases beyond {M}arkov},\ }\href {https://doi.org/10.1103/PhysRevLett.133.140403} {\bibfield  {journal} {\bibinfo  {journal} {Phys. Rev. Lett.}\ }\textbf {\bibinfo {volume} {133}},\ \bibinfo {pages} {140403} (\bibinfo {year} {2024})}\BibitemShut {NoStop}%
\bibitem [{\citenamefont {Haken}\ \emph {et~al.}(1967)\citenamefont {Haken}, \citenamefont {Risken},\ and\ \citenamefont {Weidlich}}]{haken1967quantum}%
  \BibitemOpen
  \bibfield  {author} {\bibinfo {author} {\bibfnamefont {H.}~\bibnamefont {Haken}}, \bibinfo {author} {\bibfnamefont {H.}~\bibnamefont {Risken}},\ and\ \bibinfo {author} {\bibfnamefont {W.}~\bibnamefont {Weidlich}},\ }\bibfield  {title} {\bibinfo {title} {{Quantum mechanical solutions of the laser masterequation: III. Exact equation for a distribution function of macroscopic variables}},\ }\href {https://doi.org/10.1007/BF01326496} {\bibfield  {journal} {\bibinfo  {journal} {Z. Physik}\ }\textbf {\bibinfo {volume} {206}},\ \bibinfo {pages} {355} (\bibinfo {year} {1967})}\BibitemShut {NoStop}%
\bibitem [{\citenamefont {Carmichael}(1986)}]{carhmichael1986quantum}%
  \BibitemOpen
  \bibfield  {author} {\bibinfo {author} {\bibfnamefont {H.~J.}\ \bibnamefont {Carmichael}},\ }\bibfield  {title} {\bibinfo {title} {Quantum fluctuations in absorptive bistability without adiabatic elimination},\ }\href {https://doi.org/10.1103/PhysRevA.33.3262} {\bibfield  {journal} {\bibinfo  {journal} {Phys. Rev. A}\ }\textbf {\bibinfo {volume} {33}},\ \bibinfo {pages} {3262} (\bibinfo {year} {1986})}\BibitemShut {NoStop}%
\bibitem [{\citenamefont {Carmichael}(2008)}]{carmichael2008statistical2}%
  \BibitemOpen
  \bibfield  {author} {\bibinfo {author} {\bibfnamefont {H.~J.}\ \bibnamefont {Carmichael}},\ }\href {https://doi.org/10.1007/978-3-540-71320-3} {\emph {\bibinfo {title} {{Statistical Methods in Quantum Optics 2}}}}\ (\bibinfo  {publisher} {Springer},\ \bibinfo {address} {Berlin},\ \bibinfo {year} {2008})\BibitemShut {NoStop}%
\bibitem [{\citenamefont {Annby-Andersson}\ \emph {et~al.}(2022)\citenamefont {Annby-Andersson}, \citenamefont {Bakhshinezhad}, \citenamefont {Bhattacharyya}, \citenamefont {De~Sousa}, \citenamefont {Jarzynski}, \citenamefont {Samuelsson},\ and\ \citenamefont {Potts}}]{andersson2022quantum}%
  \BibitemOpen
  \bibfield  {author} {\bibinfo {author} {\bibfnamefont {B.}~\bibnamefont {Annby-Andersson}}, \bibinfo {author} {\bibfnamefont {F.}~\bibnamefont {Bakhshinezhad}}, \bibinfo {author} {\bibfnamefont {D.}~\bibnamefont {Bhattacharyya}}, \bibinfo {author} {\bibfnamefont {G.}~\bibnamefont {De~Sousa}}, \bibinfo {author} {\bibfnamefont {C.}~\bibnamefont {Jarzynski}}, \bibinfo {author} {\bibfnamefont {P.}~\bibnamefont {Samuelsson}},\ and\ \bibinfo {author} {\bibfnamefont {P.~P.}\ \bibnamefont {Potts}},\ }\bibfield  {title} {\bibinfo {title} {Quantum {F}okker-{P}lanck master equation for continuous feedback control},\ }\href {https://doi.org/10.1103/PhysRevLett.129.050401} {\bibfield  {journal} {\bibinfo  {journal} {Phys. Rev. Lett.}\ }\textbf {\bibinfo {volume} {129}},\ \bibinfo {pages} {050401} (\bibinfo {year} {2022})}\BibitemShut {NoStop}%
\bibitem [{\citenamefont {Ptaszy\'{n}ski}\ \emph {et~al.}()\citenamefont {Ptaszy\'{n}ski}, \citenamefont {Chudak},\ and\ \citenamefont {Esposito}}]{zenodo}%
  \BibitemOpen
  \bibfield  {author} {\bibinfo {author} {\bibfnamefont {K.}~\bibnamefont {Ptaszy\'{n}ski}}, \bibinfo {author} {\bibfnamefont {M.}~\bibnamefont {Chudak}},\ and\ \bibinfo {author} {\bibfnamefont {M.}~\bibnamefont {Esposito}},\ }\href@noop {} {\bibinfo {title} {Quantum instanton theory for metastable collective spin systems}},\ \bibinfo {note} {{Z}enodo, doi: \href{https://doi.org/10.5281/zenodo.19565928}{10.5281/zenodo.19565928}}\BibitemShut {NoStop}%
\bibitem [{\citenamefont {Chudak}()}]{github}%
  \BibitemOpen
  \bibfield  {author} {\bibinfo {author} {\bibfnamefont {M.}~\bibnamefont {Chudak}},\ }\href@noop {} {\bibinfo {title} {quantum-instantons-in-collective-spin-systems}},\ \bibinfo {note} {{GitHub repository (2026), \url{https://github.com/mch-ifm/quantum-instantons-in-collective-spin-systems}}}\BibitemShut {NoStop}%
\bibitem [{\citenamefont {Alicki}\ and\ \citenamefont {Messer}(1983)}]{alicki1983nonlinear}%
  \BibitemOpen
  \bibfield  {author} {\bibinfo {author} {\bibfnamefont {R.}~\bibnamefont {Alicki}}\ and\ \bibinfo {author} {\bibfnamefont {J.}~\bibnamefont {Messer}},\ }\bibfield  {title} {\bibinfo {title} {Nonlinear quantum dynamical semigroups for many-body open systems},\ }\href {https://doi.org/10.1007/BF01012712} {\bibfield  {journal} {\bibinfo  {journal} {J. Stat. Mech.}\ }\textbf {\bibinfo {volume} {32}},\ \bibinfo {pages} {299} (\bibinfo {year} {1983})}\BibitemShut {NoStop}%
\bibitem [{\citenamefont {Benatti}\ \emph {et~al.}(2016)\citenamefont {Benatti}, \citenamefont {Carollo}, \citenamefont {Floreanini},\ and\ \citenamefont {Narnhofer}}]{benatti2016non}%
  \BibitemOpen
  \bibfield  {author} {\bibinfo {author} {\bibfnamefont {F.}~\bibnamefont {Benatti}}, \bibinfo {author} {\bibfnamefont {F.}~\bibnamefont {Carollo}}, \bibinfo {author} {\bibfnamefont {R.}~\bibnamefont {Floreanini}},\ and\ \bibinfo {author} {\bibfnamefont {H.}~\bibnamefont {Narnhofer}},\ }\bibfield  {title} {\bibinfo {title} {Non-markovian mesoscopic dissipative dynamics of open quantum spin chains},\ }\href {https://doi.org/https://doi.org/10.1016/j.physleta.2015.10.062} {\bibfield  {journal} {\bibinfo  {journal} {Phys. Lett. A}\ }\textbf {\bibinfo {volume} {380}},\ \bibinfo {pages} {381} (\bibinfo {year} {2016})}\BibitemShut {NoStop}%
\bibitem [{\citenamefont {Benatti}\ \emph {et~al.}(2018)\citenamefont {Benatti}, \citenamefont {Carollo}, \citenamefont {Floreanini},\ and\ \citenamefont {Narnhofer}}]{benatti2018quantum}%
  \BibitemOpen
  \bibfield  {author} {\bibinfo {author} {\bibfnamefont {F.}~\bibnamefont {Benatti}}, \bibinfo {author} {\bibfnamefont {F.}~\bibnamefont {Carollo}}, \bibinfo {author} {\bibfnamefont {R.}~\bibnamefont {Floreanini}},\ and\ \bibinfo {author} {\bibfnamefont {H.}~\bibnamefont {Narnhofer}},\ }\bibfield  {title} {\bibinfo {title} {Quantum spin chain dissipative mean-field dynamics},\ }\href {https://doi.org/10.1088/1751-8121/aacbdb} {\bibfield  {journal} {\bibinfo  {journal} {J. Phys. A: Math. Theor.}\ }\textbf {\bibinfo {volume} {51}},\ \bibinfo {pages} {325001} (\bibinfo {year} {2018})}\BibitemShut {NoStop}%
\bibitem [{\citenamefont {Fiorelli}\ \emph {et~al.}(2023)\citenamefont {Fiorelli}, \citenamefont {M{\"u}ller}, \citenamefont {Lesanovsky},\ and\ \citenamefont {Carollo}}]{fiorelli2023mean}%
  \BibitemOpen
  \bibfield  {author} {\bibinfo {author} {\bibfnamefont {E.}~\bibnamefont {Fiorelli}}, \bibinfo {author} {\bibfnamefont {M.}~\bibnamefont {M{\"u}ller}}, \bibinfo {author} {\bibfnamefont {I.}~\bibnamefont {Lesanovsky}},\ and\ \bibinfo {author} {\bibfnamefont {F.}~\bibnamefont {Carollo}},\ }\bibfield  {title} {\bibinfo {title} {Mean-field dynamics of open quantum systems with collective operator-valued rates: validity and application},\ }\href {https://doi.org/10.1088/1367-2630/ace470} {\bibfield  {journal} {\bibinfo  {journal} {New J. Phys.}\ }\textbf {\bibinfo {volume} {25}},\ \bibinfo {pages} {083010} (\bibinfo {year} {2023})}\BibitemShut {NoStop}%
\bibitem [{\citenamefont {Nation}\ \emph {et~al.}(2015)\citenamefont {Nation}, \citenamefont {Johansson}, \citenamefont {Blencowe},\ and\ \citenamefont {Rimberg}}]{nation2015iterative}%
  \BibitemOpen
  \bibfield  {author} {\bibinfo {author} {\bibfnamefont {P.~D.}\ \bibnamefont {Nation}}, \bibinfo {author} {\bibfnamefont {J.~R.}\ \bibnamefont {Johansson}}, \bibinfo {author} {\bibfnamefont {M.~P.}\ \bibnamefont {Blencowe}},\ and\ \bibinfo {author} {\bibfnamefont {A.~J.}\ \bibnamefont {Rimberg}},\ }\bibfield  {title} {\bibinfo {title} {Iterative solutions to the steady-state density matrix for optomechanical systems},\ }\href {https://doi.org/10.1103/PhysRevE.91.013307} {\bibfield  {journal} {\bibinfo  {journal} {Phys. Rev. E}\ }\textbf {\bibinfo {volume} {91}},\ \bibinfo {pages} {013307} (\bibinfo {year} {2015})}\BibitemShut {NoStop}%
\bibitem [{\citenamefont {Dubois}\ \emph {et~al.}(2021)\citenamefont {Dubois}, \citenamefont {Saalmann},\ and\ \citenamefont {Rost}}]{dubois2021semi}%
  \BibitemOpen
  \bibfield  {author} {\bibinfo {author} {\bibfnamefont {J.}~\bibnamefont {Dubois}}, \bibinfo {author} {\bibfnamefont {U.}~\bibnamefont {Saalmann}},\ and\ \bibinfo {author} {\bibfnamefont {J.~M.}\ \bibnamefont {Rost}},\ }\bibfield  {title} {\bibinfo {title} {Semi-classical {L}indblad master equation for spin dynamics},\ }\href {https://doi.org/https://iopscience.iop.org/article/10.1088/1751-8121/abf79b/meta} {\bibfield  {journal} {\bibinfo  {journal} {J. Phys. A: Math. Theor.}\ }\textbf {\bibinfo {volume} {54}},\ \bibinfo {pages} {235201} (\bibinfo {year} {2021})}\BibitemShut {NoStop}%
\end{thebibliography}%

\end{document}


	
	\title{Supplemental Material to ``Quantum instanton theory for metastable collective spin systems''}
	
	\author{Krzysztof Ptaszy\'{n}ski}
	\email{krzysztof.ptaszynski@ifmpan.edu.pl}
 	\affiliation{Institute of Molecular Physics, Polish Academy of Sciences, Mariana Smoluchowskiego 17, 60-179 Pozna\'{n}, Poland}

\author{Maciej Chudak}
 	\affiliation{Institute of Molecular Physics, Polish Academy of Sciences, Mariana Smoluchowskiego 17, 60-179 Pozna\'{n}, Poland}
    
	\author{Massimiliano Esposito}
\affiliation{Complex Systems and Statistical Mechanics, Department of Physics and Materials Science, University of Luxembourg, 30 Avenue des Hauts-Fourneaux, L-4362 Esch-sur-Alzette, Luxembourg}
	
	\date{\today}
	
	\maketitle
This Supplemental Material contains the following sections:
    \begin{itemize}
    \item Details of the quantum master equation (QME) approach (Sec.~\ref{supp:qme})
    \item Description of the numerical continuation method (Sec.~\ref{supp:cont})
    \item Description of the SW approach (Sec.~\ref{supp:sw})
    \item Results for $\Omega=0.5\gamma$ (Sec.~\ref{supp:05gamma})
    \item Derivations of Hamiltonians $\mathcal{H}_\alpha$ (Sec.~\ref{supp:derhalpha})
    \end{itemize}

    \section{Master equation} \label{supp:qme}

    Here we describe the formalism used to find the steady state of the QME and the Liouvillian gap $\lambda$. To that end, we represent the density matrix in terms of eigenstates of $\hat{J}_z|J,M\rangle=M|J,M\rangle$ as $\hat{\rho}=\sum_{M,M'=-J}^J \rho_{MM'}|J,M\rangle\langle J,M'|$. The matrix elements then evolve as
\begin{align} \label{eq:masteq-orig} \nonumber
& d_t \rho_{M,M'}=-\frac{i \Omega}{2} \left(C_{M-1} \rho_{M-1,M'} +C_M \rho_{M+1,M'} \right. \\ \nonumber &\left. -C_{M'-1} \rho_{M,M'-1} - C_{M'} \rho_{M,M'+1} \right) \\ \nonumber
&+ \frac{\gamma}{J} \left[C_{-M} C_{-M'} \rho_{M-1,M'-1}-\frac{1}{2} \left(C_M^2 +C_{M'}^2 \right) \rho_{M,M'} \right] \\
&\nonumber + \frac{\Gamma}{J^3} \Big[(M+1)(M'+1)C_{M} C_{M'} \rho_{M+1,M'+1} \\ & -\frac{1}{2} \left(M^2C_{-M}^2 +M'^2 C_{-M'}^2 \right) \rho_{M,M'} \Big] \,,
\end{align}
where $C_M=\sqrt{J(J+1)-M(M+1)}$. We now note that---due to the system's mirror symmetry with respect to $(y,z)$ plane---one can formulate a closed set of equations for real elements $p_{M,M+k}$ with
$\rho_{M,M+k}=p_{M,M+k}$ for even $k$ and $\rho_{M,M+k}=ip_{M,M+k}$ for odd $k$. It reads as
\begin{subequations}
\begin{align} \nonumber
&d_t p_{M,M+k}=\frac{(-1)^k\Omega}{2} \left(C_{M-1} p_{M-1,M+k} +C_M p_{M+1,M+k} \right.\\ & \nonumber \left. -C_{M+k-1} p_{M,M+k-1} - C_{M+k} p_{M,M+k+1} \right) \\ \nonumber
&+ \frac{\gamma}{J} \left[C_{-M} C_{-M-k} p_{M-1,M+k-1}-\frac{1}{2} \left(C_M^2 +C_{M+k}^2 \right) p_{M,M+k} \right] \\
&\nonumber + \frac{\Gamma}{J^3} \Big\{(M+1)(M+k+1)C_{M} C_{M+k} p_{M+1,M+k+1} \\ &-\frac{1}{2} \left[M^2C_{-M}^2 +(M+k)^2 C_{-M-k}^2 \right] p_{M,M+k} \Big\} \quad \text{for $k>0$} \,,  \\ \nonumber
&d_t p_{M,M}=\Omega \left(C_{M-1} p_{M-1,M} -C_M p_{M,M+1} \right) \\ \nonumber &+ \frac{\gamma}{J} \left(C_{-M}^2 p_{M-1,M-1}-C_M^2 p_{M,M} \right) \\&+ \frac{\Gamma}{J^3} \left[(M+1)^2 C_{M}^2 p_{M+1,M+1}-M^2C_{-M}^2 p_{M,M} \right] \quad \text{for $k=0$} \,,
\end{align}
\end{subequations}
where all coefficients are real and where the number of independent variables $p_{M,M+k}$ is reduced from $(2J+1)^2$ to $(2J+1)(J+1)$. Physically, this corresponds to the evolution of the part of the density matrix $\hat{\rho}$ that is symmetric with respect to the $(y,z)$ plane, which is decoupled from the dynamics of the antisymmetric part. The expressions above can be rewritten as a matrix equation 
\begin{align}
d_t\boldsymbol{p}=\mathbb{W} \boldsymbol{p} \,,
\end{align} 
where $\boldsymbol{p}$ is the column vector of $p_{M,M+k}$. The steady state is determined by solving the linear equation~\cite{nation2015iterative}
\begin{align} \label{eq:pss} (\mathbb{W}+g\boldsymbol{e}\boldsymbol{1}^\intercal) \boldsymbol{p}=g\boldsymbol{e} \,,
\end{align}
where $g$ is an arbitrary weighting factor, while $\boldsymbol{e}$ ($\boldsymbol{1}$) is the column vector with one (all) elements at positions corresponding to $p_{M,M}$ equal to 1, and other elements 0. This is formally equivalent to  solving $\mathbb{W} \boldsymbol{p}=0$ with the normalization condition $\sum_{M=-J}^J p_{MM}=1$, but it is more numerically robust. Equation~\eqref{eq:pss} has been solved using the LinearSolve function in Wolfram Mathematica, with the GMRES method and ILU0 preconditioner. Due to the large condition number (related to the small Liouvillian gap $\lambda$), we used an increased numerical precision of 50 digits. Magnetization $m_{z,\text{ss}}$ has been calculated as
\begin{align}
m_{z,\text{ss}}=\sum_{M=-J}^J M p_{M,M}/J \,.
\end{align}

Since the slowest relaxation timescale of the model is related to relaxation in the symmetric sector of $\hat{\rho}$, the Liouvillian gap can be determined as a spectral gap of the matrix $\mathbb{W}$. We did it using the Eigenvalues function in Wolfram Mathematica employing the Arnoldi algorithm. Equivalently, it can be determined using the matrix representation of the original equation~\eqref{eq:masteq-orig}, but this is more numerically demanding. 

\section{Numerical continuation method} \label{supp:cont}
To determine instanton trajectories, we note that the Hamiltonian $\mathcal{H}_\alpha$ for $v=\pi_v=0$ can be expressed as
\begin{align} \label{eq:zeroencond}
\mathcal{H}_\alpha(w,\pi_w)=\pi_w \mathcal{C}_\alpha (w,\pi_w) \,,
\end{align}
where $\mathcal{C}_\alpha$ is the cubic polynomial
\begin{align}
\mathcal{C}_\alpha (w,\pi_w)=\sum_{j=0}^3 \pi_w^j \mathcal{C}_\alpha^{(j)} \,,
\end{align}
with the coefficients
\begin{subequations}
    \begin{align}
&\mathcal{C}^{(0)}_H=\mathcal{C}^{(0)}_P=\frac{\Omega}{2} (1+w^2)+\gamma w-\Gamma w \frac{(w^2-1)^2}{(w^2+1)^2} \,, \\
&\mathcal{C}^{(1)}_H=\frac{\gamma  \left(w^3+w\right)^2+\Gamma  \left(5 w^4-6 w^2+1\right)}{4 \left(w^2+1\right)} \,, \\
&\mathcal{C}^{(2)}_H=\frac{\Gamma}{4} \left(  w-2  w^3\right) \,, \\
&\mathcal{C}^{(3)}_H=\frac{\Gamma w^2}{16} \left(w^2+1\right) \,, \\
&\mathcal{C}^{(1)}_P=\frac{\gamma  \left(w^2+1\right)^2+\Gamma  \left(w^4-6 w^2+5\right) w^2}{4 \left(w^2+1\right)} \,, \\
&\mathcal{C}^{(2)}_P= \frac{\Gamma  w^3}{4} \left(w^2-2\right) \,, \\
&\mathcal{C}^{(3)}_P=\frac{\Gamma w^4}{16} \left(w^2+1\right) \,.
\end{align}
\end{subequations}
We now parametrize the instanton trajectory by its arclength $s$ as $[w(s),\pi_w(s)]$. Since $\mathcal{H}_\alpha=0$ along the instanton while $\pi_w(s)\neq 0$, by Eq.~\eqref{eq:zeroencond} the trajectory satisfies
\begin{align}
\mathcal{C}_\alpha[w(s),\pi_w(s)]=0\,.
\end{align}
The trajectory satisfying this condition is constructed by numerical continuation. To determine the instanton segment connecting the stable FP $\boldsymbol{x}_i^*=(0,w_i^*)$ to the unstable FP $\boldsymbol{x}_\nu^*=(0,w_\nu^*)$, where $\nu\in{s_1,r_1}$, we propagate the differential equations
\begin{subequations} \label{eq:contdif}
    \begin{align}
d_s w&=\frac{\pm \partial_{\pi_w} \mathcal{C}_\alpha-\tau \mathcal{C}_\alpha \partial_{w} \mathcal{C}_\alpha}{\sqrt{(\partial_{w} \mathcal{C}_\alpha)^2+(\partial_{\pi_w} \mathcal{C}_\alpha)^2}} \,, \\
d_s \pi_w&=\frac{\mp \partial_w \mathcal{C}_\alpha-\tau \mathcal{C}_\alpha \partial_{\pi_w} \mathcal{C}_\alpha}{\sqrt{(\partial_{w} \mathcal{C}_\alpha)^2+(\partial_{\pi_w} \mathcal{C}_\alpha)^2}} \,,
\end{align}
\end{subequations}
with the initial point $[w(0),\pi_w(0)]=(w_i^*,0)$. These equations propagate the trajectory along the vector perpendicular to the gradient $(\partial_w \mathcal{C}_\alpha,\partial_{\pi_w} \mathcal{C}_\alpha)$. The sign $\pm$ corresponds to two distinct instanton trajectories leaving $\boldsymbol{x}_i^*$. The terms proportional to an arbitrary parameter $\tau>0$ have been added to improve numerical stability---they relax any deviations of the trajectory from the $\mathcal{C}_\alpha=0$ manifold toward that manifold. However, the actual value of $\tau$ does not significantly affect the results and no significant change is observed even when $\tau = 0$.

Equations~\eqref{eq:contdif} are propagated up to the point $s_\nu^*$ where $\pi_w(s_\nu^*)=0$ and thus $w(s_\nu^*)=w^*_\nu$. The action along the trajectory $\boldsymbol{x}_i^* \rightarrow \boldsymbol{x}^*_\nu$ can then be calculated by converting Eq.~(11) from the main text into integration over the arclength coordinate~\cite{zakine2023minimum},
\begin{align}
S_{\alpha}(\boldsymbol{x}^*_\nu,t|\boldsymbol{x}_i^*,0)=\int_{0}^{s_\nu^*} ds \pi_w(s) d_s w(s) \,.
\end{align}

\section{SW approach} \label{supp:sw}
Here, we review the semiclassical Wigner (SW) approach developed in Ref.~\cite{dubois2021semi} against which we compare our framework. Within this method, the action of the Liouvillian superoperator
\begin{align}
\mathcal{L} \bullet\equiv -i[\hat{H},\bullet] +\sum_j \eta_j \mathcal{D}[\hat{L}_j] \bullet \,,
\end{align}
is represented by the differential operator $L_\text{SW}(\boldsymbol{x},\nabla_{\boldsymbol{x}})$ that describes the evolution of the Wigner distribution $p_W(\boldsymbol{x})$, which is truncated after the second derivative term. For our model, we have $\hat{H}=\Omega \hat{J}_x$, $\eta_1=\gamma/J$, $\hat{L}_1=\hat{J}_+$, $\eta_2=\Gamma/J^3$ and $\hat{L}_2=\hat{J}_- \hat{J}_z$. The operator $L_\text{SW}$ has Fokker--Planck form 
\begin{align}
L_\text{SW}(\boldsymbol{x},\nabla_{\boldsymbol{x}})=-\sum_k \frac{\partial}{\partial x_k} h_k(\boldsymbol{x}) + \frac{1}{J} \sum_{k,l} \frac{\partial}{\partial x_k} \frac{\partial}{\partial x_l} D_{kl}(\boldsymbol{x}) \,, 
\end{align}
with drift and diffusion terms
\begin{align} \nonumber
&h_k(\boldsymbol{x})=\{x_k,H \}_\text{PB} \\ \nonumber &-\sum_j \frac{i \eta_j}{2} \left(L_j \left\{x_k,L_j^* \right\}_\text{PB}+L_j^* \left\{L_j ,x_k\right\}_\text{PB} \right)+O(1/J)\,, \\
&D_{kl}(\boldsymbol{x})=\sum_{j} \frac{J \eta_j}{4} \big( \{x_k,L_j\}_\text{PB} \{x_l,L_j^*\}_\text{PB}+\text{c.c} \big) \,.
\end{align}
Here, the functions $H$, $L_j$ are obtained from $\hat{H}$, $\hat{L}_j$ by replacing the operators $\hat{J}_k$ with scalars $J_k$. The object
\begin{align}
\{F,G\}_\text{PB} \equiv \boldsymbol{J} \cdot \left( \nabla_{\boldsymbol{J}} F \times  \nabla_{\boldsymbol{J}} G \right) \,,
\end{align}
where $\boldsymbol{J}=(J_x,J_y,J_z)$, is the spin Poisson bracket. For the stereographic coordinates $\boldsymbol{x}=(v,w)$, we further have $v=J_x/(J-J_z)$ and $w=J_y/(J-J_z)$. The corresponding Hamiltonian $\mathcal{H}_\text{SW}$ has a quadratic form
\begin{align}
&\mathcal{H}_\text{SW}(\boldsymbol{x},\boldsymbol{\pi}) \\ \nonumber &=\pi_v h_v+\pi_w h_w+\pi_v^2 D_{vv} +\pi_w^2 D_{ww}+\pi_v \pi_w (D_{vw}+D_{wv}) \,,
\end{align}
with the drift terms
\begin{subequations}
    \begin{align}
h_v&=\Omega vw+\gamma v-\Gamma v \frac{(v^2+w^2-1)^2}{(v^2+w^2+1)^2} \,, \\
h_w&=\frac{\Omega}{2} (1-v^2+w^2)+\gamma w-\Gamma w \frac{(v^2+w^2-1)^2}{(v^2+w^2+1)^2} \,,
\end{align}
\end{subequations}
and the diffusion terms
\small
\begin{widetext}
\begin{subequations}
\begin{align}
&D_{vv}=\frac{1}{8} \left\{\gamma  \left[v^2+(w-1)^2\right] \left[v^2+(w+1)^2\right]+\frac{\Gamma  \left[v^8+4 v^6 w^2+2 v^4 \left(3 w^4-6 w^2-1\right)+4 v^2 w^2 \left(w^2-3\right)^2+\left(w^4-6 w^2+1\right)^2\right]}{\left(v^2+w^2+1\right)^2}\right\} \,, \\
&D_{ww}=\frac{1}{8} \left\{\gamma  \left[(v-1)^2+w^2\right] \left[(v+1)^2+w^2\right]+\frac{\Gamma  \left[4 v^2 w^6+4 v^2 \left(v^2-3\right)^2 w^2+2 \left(3 v^4-6 v^2-1\right) w^4+\left(v^4-6 v^2+1\right)^2+w^8\right]}{\left(v^2+w^2+1\right)^2}\right\} \,, \\
&D_{vw}=D_{wv}=\frac{1}{2} v w \left[\gamma +\frac{\Gamma  \left(v^2+w^2-3\right) \left(3 v^2+3 w^2-1\right)}{\left(v^2+w^2+1\right)^2}\right] \,.
\end{align}
\end{subequations}
\end{widetext}
\normalsize
Their derivation is presented in Wolfram Mathematica notebooks available at~\cite{zenodo} and printed in Sec.~\ref{supp:derhalpha}.

Let us now turn to the determination of activation barriers $\mathcal{A}_{i \rightarrow j}$. Since instanton trajectories are confined to the $w$-$\pi_w$ plane, they can be determined using the approach from Sec.~\ref{supp:cont}. However, the analysis becomes even simpler when noting that the reduced Hamiltonian with $v=\pi_v=0$ has the quadratic form
\begin{align}
\mathcal{H}_\text{SW}(w,\pi_w)=\pi_w h_w(w)+\pi_w^2D_{ww}(w) \,,
\end{align}
where
\begin{align} \nonumber
h_w(w) &=\frac{\Omega}{2} (1+w^2)+\gamma w-\Gamma w \frac{(w^2-1)^2}{(w^2+1)^2}\,,\\
D_{ww}(w) &=\frac{\gamma}{8}(w^2+1)^2+\frac{\Gamma}{8}(w^2-1)^2 \,.
\end{align}
Since instantons are located in the $\mathcal{H}_\text{SW}=0$ manifold, the momentum along the instanton trajectories reads $\pi_w=-h_w(w)/D_{ww}(w)$. The action along the instanton trajectory from a stable FP $\boldsymbol{x}_i^*=(0,w_i^*)$ to an unstable FP $\boldsymbol{x}_\nu^*=(0,w_\nu^*)$, $\nu \in \{s_1,r_1\}$, can then be calculated as
\begin{align}
S_\text{SW}(\boldsymbol{x}_\nu^*,t|\boldsymbol{x}_i^*,0)=-\int_{w_i^*}^{w_\nu^*} dw h_w(w)/D_{ww}(w) \,.
\end{align}

\section{Results for $\Omega=0.5 \gamma$} \label{supp:05gamma}

\begin{figure}
    \centering
    \includegraphics[width=0.9\linewidth]{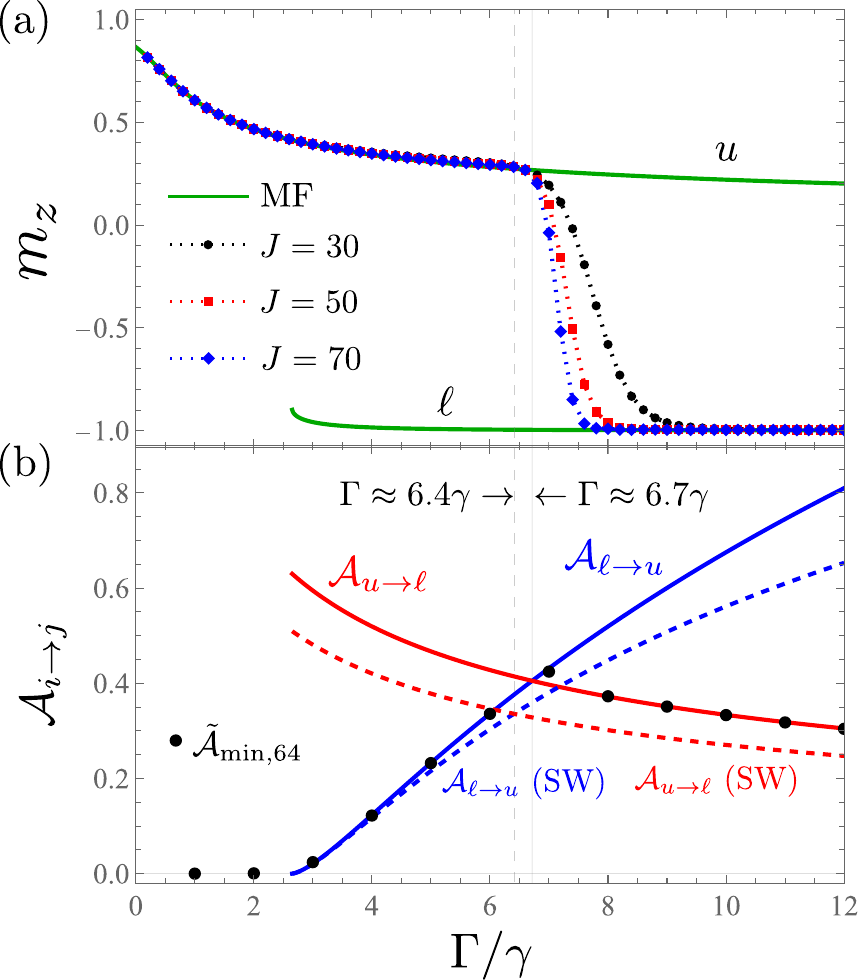}
    \caption{(a) MF fixed-point magnetization $m_z^*$ (green solid lines) versus finite-size steady-state magnetization $m_{z,\text{ss}}$ obtained using QME (symbols). (b) Activation barriers $\mathcal{A}_{i \rightarrow j}$ calculated using our approach (solid lines) versus SW approach (dashed lines). Grey vertical solid (dashed) line denotes the crossing point of $\mathcal{A}_{\ell \rightarrow u}$ and $\mathcal{A}_{u \rightarrow \ell}$ curves for our (SW) approach. Dots denote the estimator $\tilde{\mathcal{A}}_{\text{min},64}$ defined by Eq.~\eqref{eq:aminestsupp} [Eq.~(14) in the main text]. Parameter $\Omega=0.5 \gamma$.
    }
    \label{fig:mzaction-05}
\end{figure}

\begin{figure}
    \centering
    \includegraphics[width=0.9\linewidth]{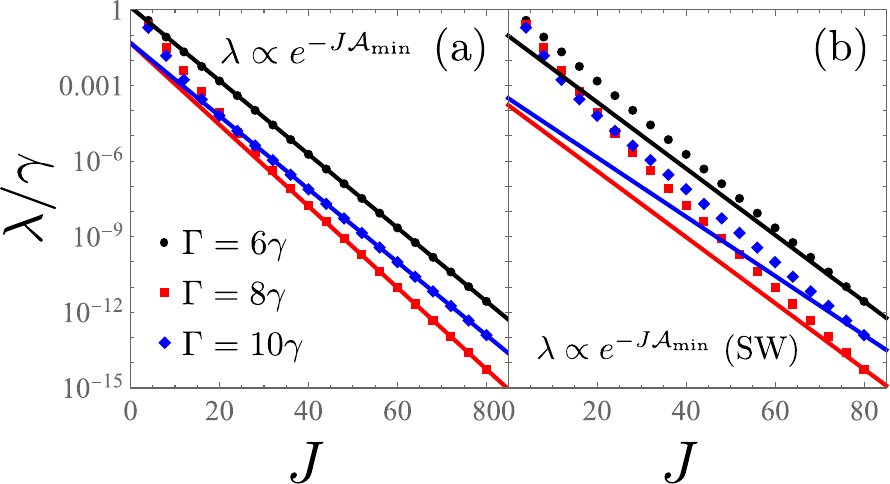}
    \caption{(a) Dots: Liouvillian gap $\lambda$ calculated using QME for different $\Gamma$, plotted in the logarithmic scale. Lines: the predicted scaling $\lambda \propto \exp(-J \mathcal{A}_\text{min})$ set to cross the point for $J=80$. (b) The same plot, but with $\mathcal{A}_\text{min}$ given by the SW approach. Parameter $\Omega=0.5 \gamma$.
    }
    \label{fig:liovgap-05}
\end{figure}

In Figs.~\ref{fig:mzaction-05} and \ref{fig:liovgap-05} we plot the equivalents of Figs.~1 and~3 from the main text, but for larger $\Omega=0.5\gamma$. Based on them, we can reach the same conclusions: our approach correctly characterizes the position of the first-order phase transition (the point where $m_{z,\text{ss}}$ crosses over from $u$ to $\ell$ branch due to crossing of $\mathcal{A}_{\ell\rightarrow u}$ and $\mathcal{A}_{u \rightarrow \ell}$) and the asymptotic scaling of the Liouvillian gap, while these features are not accurately described by the SW approach. We also note that the activation barriers $\mathcal{A}_{i \rightarrow j}$ are now approximately twice as small as for $\Omega=0.25\gamma$, and that the difference between our framework and the SW approach with respect to the position of the first-order phase transition is now three times smaller (reduced from $\approx \gamma$ to $\approx 0.3\gamma$).

Concerning the estimator $\tilde{\mathcal{A}}_{\text{min},J}$, we note that it perfectly agrees with the obtained value of $\mathcal{A}_\text{min} \equiv \min(\mathcal{A}_{\ell\rightarrow u},\mathcal{A}_{u \rightarrow \ell})$, apart from $\Gamma=7\gamma$, where a few-percent difference is observed. This results from finite-size effects: Even though for this point $\mathcal{A}_{\ell \rightarrow u}>\mathcal{A}_{u \rightarrow \ell}$, the branch $u$ is still occupied with finite probability for the considered values of $J$, as implied by the $m_{z,\text{ss}}$ plot [Fig.~\ref{fig:mzaction-05}(a)]. Consequently, the Liouvillian gap is determined by the sum of the two switching rates~\cite{carde2026nonperturvative}
\begin{align}
\lambda\approx \kappa_{\ell\rightarrow u}+\kappa_{u\rightarrow \ell}=\omega_{\ell\rightarrow u}e^{-J \mathcal{A}_{\ell \rightarrow u} } +\omega_{u\rightarrow \ell}e^{-J \mathcal{A}_{u \rightarrow \ell} } \,,
\end{align}
where $\omega_{i \rightarrow j}$ is a subexponential prefactor, $\lim_{J \rightarrow \infty} (\ln \omega_{i \rightarrow j})/J=0$. For finite $J$, both terms therefore contribute to the gap, so that the estimator  [see Eq.~(14) in the main text]
\begin{align} \label{eq:aminestsupp}
\tilde{\mathcal{A}}_{\text{min},J} \equiv \frac{1}{4} \ln \frac{\lambda(J-4)}{\lambda(J)} \,
\end{align}
assumes an effective value between $\mathcal{A}_{\ell\rightarrow u}$ and $\mathcal{A}_{u\rightarrow\ell}$, as observed in Fig.~\ref{fig:mzaction-05}(b).

From a technical point of view, we note that $m_{z,\text{ss}}$ has been determined now using the Multifrontal method with machine precision, rather than the GMRES method. This is because this approach was more numerically stable, while for $\Omega=0.25\gamma$ the situation was the opposite.

\section{Derivations of $\mathcal{H}_\alpha$} \label{supp:derhalpha}
Below the bibliography, we copy the prints of the Wolfram Mathematica notebooks used to derive Hamiltonians $\mathcal{H}_\alpha$. This not only illustrates the final result of the derivations, but also shows how these derivations can be easily performed using computational algebra software. The notebooks themselves are available at~\cite{zenodo}.

\bibliography{bibliography}

\clearpage
\foreach \p in {1,2,3,4,5,6,7,8} { 
    \begin{figure}[p] 
        \centering
        \vspace*{-2cm} 
        \includegraphics[page=\p, width=0.9\paperwidth, angle=0]{derivation-combined.pdf}
    \end{figure}
    \clearpage }